\documentclass[12pt,english]{article}
\usepackage[latin1]{inputenc}
\usepackage{color}
\makeatletter

\newcommand{\be}{\begin{equation}}
\newcommand{\ee}{\end{equation}}
\newcommand{\bea}{\begin{eqnarray}}
\newcommand{\eea}{\end{eqnarray}}
\newcommand{\ba}{\begin{array}}
\newcommand{\ea}{\end{array}}

\def\bbox{{\,\lower0.9pt\vbox{\hrule \hbox{\vrule height 0.2 cm
\hskip 0.2 cm \vrule height 0.2 cm}\hrule}\,}}
\newcommand{\dsl}{\pa \kern-0.5em /}

\usepackage{latexsym}
\usepackage{amsmath,,calrsfs}
\usepackage{amsfonts}
\usepackage{amssymb}
\usepackage{amscd}
\usepackage{bbm}
\usepackage{fancybox}
\usepackage{cite}
\usepackage{amsmath,amsfonts,amsbsy}
\usepackage{pstricks,pst-node}
\usepackage[small,bf,hang]{caption2}
\usepackage{graphicx}
\usepackage{epsfig}
\usepackage{psfrag}
\usepackage{comment}

\font\mybb=msbm10 at 12pt
\def\bb#1{\hbox{\mybb#1}}
\def\bZ {\bb{Z}}
\def\bR {\bb{R}}

\def\bC {\bb{C}}

\font\mybb=msbm10 at 12pt
\def\bb#1{\hbox{\mybb#1}}

\def\bX {\bb{X}}
\def\bP {\bb{P}}

\def\bU {\bb{U}}
\def\bV {\bb{V}}
\def\bI {\bb{I}}

\usepackage{babel}
\makeatother
\begin{document}



\begin{titlepage}
\rightline{UMTG-23, DAMTP-2011-24}

\vfill

\begin{center}
\baselineskip=16pt {\Large\bf  Quantum 3D  Superstrings}
\vskip 0.3cm
{\large {\sl }}
\vskip 10.mm
{\bf Luca Mezincescu$^{\dagger,1}$ and  Paul K.
Townsend$^{+,2}$}
\vskip 1cm
{\small
$^\dagger$
Department of Physics, University of Miami,\\
Coral Gables, FL 33124, USA\\
}
\vspace{6pt}
{\small
  $^+$
Department of Applied Mathematics and Theoretical Physics\\
Centre for Mathematical Sciences, University of Cambridge\\
Wilberforce Road, Cambridge, CB3 0WA, UK\\
}
\end{center}
\vfill

\par
\begin{center}
{\bf ABSTRACT}
\end{center}
\begin{quote}

The classical Green-Schwarz superstring action,  with ${\cal N}=1$ or ${\cal N}=2$ spacetime supersymmetry,  exists for spacetime dimensions $D=3,4,6,10$, but quantization in the light-cone gauge breaks  Lorentz invariance unless {\it either}  $D=10$, which leads to 
critical superstring theory, {\it or} $D=3$. We give details of results presented previously for  the bosonic and ${\cal N}=1$ closed 3D (super)strings and extend them to  the ${\cal N}=2$ 3D superstring. In all cases, the spectrum is parity-invariant  and contains anyons of irrational spin.

\vfill
\vfill
\vfill
\vfill
  \hrule width 5.cm
\vskip 2.mm
{\small
\noindent $^1$ mezincescu@server.physics.miami.edu\\
\noindent $^2$ p.k.townsend@damtp.cam.ac.uk
\\ }
\end{quote}
\end{titlepage}

\section{Introduction}
\setcounter{equation}{0}

Quantization of a relativistic string in a $D$-dimensional Minkowski background spacetime is problematic unless $D$ is the critical dimension  ($D=26$ for the Nambu-Goto string and $D=10$ for superstrings). The difficulty is  seen most clearly in the light-cone gauge; unitarity is then manifest, as all unphysical `gauge' modes  of the string are  absent, but quantum anomalies break Lorentz invariance in any (generic) non-critical dimension \cite{Goddard:1973qh} (see also  \cite{Green:1987sp},  and  \cite{Bering:2011uf} for a recent detailed computation). A corollary  is that  Lorentz-covariant quantization in a (generic) non-critical dimension can lead to a unitary theory only if it  involves some ``longitudinal'' modes\footnote{In the Nambu-Goto formulation, some such modes have a classical interpretation \cite{Bardeen:1975gx}, but we postpone discussion of this point.}. A corollary  is that  Lorentz-covariant quantization in a (generic) non-critical dimension can lead to a unitary theory only if it  involves some  additional ``longitudinal'' mode, e.g. a Liouville mode. In fact, this option is available only in sub-critical dimensions and it  has not yet proved useful for $D>2$ (see e.g. \cite{Polchinski:1998rq}).

These problems with non-critical string theories are well-known except for the  qualification  ``generic'', which refers 
to an exception that we exploited  in an earlier paper  \cite{Mezincescu:2010yp}  to which the present paper is a sequel: light-cone gauge quantization preserves Lorentz invariance not only in the critical dimension but also for $D=3$ (3D), trivially for the Nambu-Goto string\footnote{This was pointed out  at the May 2010 Solvay workshop on ``Symmetries and Dualities in Gravitational Theories'' in a talk by one of us based on a draft  version of our   subsequent paper, and also by T. Curtright  in  independent work on a related topic over the same period  \cite{Curtright:2010zz}. We have been led to understand that the  exceptional status of the bosonic 3D string was already known to experts but we are not aware of any earlier reference. Some classical aspects of the light-cone gauge for  3D strings have been discussed previously by Siegel \cite{Siegel:1983ke}.}. The light-cone gauge quantization of the  3D Nambu-Goto closed string was carried out in \cite{Mezincescu:2010yp} and it was confirmed that  Lorentz invariance is preserved in the quantum theory, without the need for any ``longitudinal'' modes.  It was also noted in \cite{Mezincescu:2010yp} that the low-lying states of non-zero spin appear in parity doublets.  Here we prove that this was no coincidence: the quantum theory preserves parity as well as Lorentz invariance. 

The 3D Nambu-Goto closed string is sufficiently simple that one can easily  determine the Lorentz representations of the states in low-lying levels explicitly  (rather than having to rely on implicit arguments based on matching degeneracies to dimensions of Lorentz representations).  The spin of the states in  levels  2 and 3 was found to depend on the intercept parameter, not surprisingly but there is no choice of this parameter for which the spins in both these levels are either integral or half-integral; in other words, the spectrum contains anyons\footnote{By ``anyon''  we mean a particle with spin not equal to an integer or half-integer. This differs, in principle,  from the definition in terms of statistics but spin and statistics are related by the 3D spin-statistics theorem; see e.g. \cite{Mund:2008hn}.}. 

We also observed in   \cite{Mezincescu:2010yp} that the spectrum contains {\it irrational} spins for a generic allowed choice of the intercept parameter.  Here we further show, by computation of the spectrum at level 4,  that  some states {\it necessarily}  have irrational spin. This  result is significant because it implies that the Lorentz group of the quantum 3D string is neither $SO(2,1)$ nor any finite multiple cover, such as the double cover  $Sl(2;\bC)$, but rather its universal cover  $\overline{SO}(2,1)$.  Irrational spin irreps of $\overline{SO}(2,1)$ are infinite-dimensional \cite{Jackiw:1990ka,Plyushchay:1990rt}, so an infinite component field is needed for any manifestly  Lorentz-invariant field theoretic description  of a particle of irrational spin.  Since irrational spin particles appear in the 3D string spectrum, it should not be a surprise that the Lorentz invariance of 3D quantum strings cannot easily be seen using current methods of covariant quantization.

The Nambu-Goto string has a natural generalization to a  spacetime supersymmetric Green-Schwarz  (GS) superstring, which exists classically for spacetime dimension $D=3,4,6,10$, with either ${\cal N}=1$ or ${\cal N}=2$ supersymmetry \cite{Green:1983wt}.   The GS superstring 
action has a fermionic ``$\kappa$-symmetry'' gauge invariance, in addition to worldsheet reparametrization invariance,  but there  is an extension of the light-cone gauge that again eliminates all ``longitudinal'' modes. 
Quantization in $D=10$ leads to standard critical superstring theory (after the inclusion of open strings in the ${\cal N}=1$ case). Light-cone gauge quantization of the  3D ${\cal N}=1$ GS superstring was carried out in \cite{Mezincescu:2010yp}.  Not only are there no Lorentz anomalies but there are also no super-Poincar\'e anomalies, and parity is also preserved. 
Moreover, the intercept parameter is now fixed by supersymmetry such that the  ground state, which is doubly degenerate,  is massless. This is entirely consistent with the possibility that there exists a  3D ${\cal N}=1$ superstring theory with an effective ${\cal N}=1$ 3D supergravity action since  only the dilaton and dilatino of the latter would propagate massless modes. 

It is convenient to refer to the Nambu-Goto string with zero intercept parameter as the ${\cal N}=0$ string; this string has spin-$3/2$ states at level-$2$ and irrational spin anyons at level-$3$. The ${\cal N}=1$ string is  a  3D heterotic string in the sense that its spectrum is a tensor product of Lorentz irreps arising from an ${\cal N}=0$ right-moving sector  with supermultiplets from an ${\cal N}=1$ supersymmetric left-moving sector. The spectrum was computed through level-$2$ in \cite{Mezincescu:2010yp}:   it was found that there are semion states (spin $1/4 + n/2$ for integer $n$) at levels $1$ and $2$ (and  irrational spins must occur at higher levels because they are present for ${\cal N}=0$).  By tensoring two  factors of the supersymmetric left-moving sector of the ${\cal N}=1$ string, one can deduce from the results of \cite{Mezincescu:2010yp} that the  ${\cal N}=2$ string has only bosons and fermions through level-$2$,  
and there is no obvious reason why anyons should appear in higher levels. This is one reason why a discussion of the ${\cal N}=2$ superstring was omitted (aside from a comment about zero-mass modes) from \cite{Mezincescu:2010yp}: it was not clear that it 
exemplified  our title ``Anyons from Strings''. 

The principal purpose of this paper is  to extend the results of  \cite{Mezincescu:2010yp} to the 3D ${\cal N}=2$ GS superstring, but we also present details, omitted from the very brief account in \cite{Mezincescu:2010yp},  of the quantization of the 3D Nambu-Goto string and of the 3D ${\cal N}=1$ GS superstring; in all cases, we shall restrict our attention to closed oriented strings. The main issue that we wish to address for ${\cal N}=2$ is whether the spectrum contains anyons. If not then we would need to explain why this quantum 3D string  had  not previously been found using  Lorentz-covariant quantization methods. It might have been necessary to invoke the usual difficulties with $\kappa$-symmetry, but a computation of the spectrum at level-$3$ suffices to show that irrational spins are also present in the spectrum of the  ${\cal N}=2$ 3D GS superstring. 

We begin with  a preliminary section that recalls pertinent features  of 3D physics and introduces some of our notation. A novelty of this section is a `re-interpretation' of the 3D supersymmetry algebra as the algebra of a model of supersymmetric quantum mechanics. This simplifies the analysis  of  the structure of  massive 3D supermultiplets. 

We then consider, in succession, the 3D closed Nambu-Goto  string, the ${\cal N}=1$ GS superstring, and finally  the ${\cal N}=2$ GS superstring. In each case  we show how gauge invariances may be fixed so as to leave only the residual global gauge invariance under shifts of the string coordinate $\sigma$, which becomes the level-matching condition in the quantum theory.  In this we follow the classic work of Goddard et al. \cite{Goddard:1973qh} except that we start with the Hamiltonian form of the string action and thus obtain directly the Hamiltonian form of the light-cone gauge-fixed action; this simplifies the verification of  (super)Poincar\'e invariance of  the gauge-fixed quantum (super)string. Having established (super)Poincar\'e invariance, we then compute the spectrum at the  first few levels, sufficient to show that the spectrum of each of the quantum (super)strings considered contains anyons of irrational spin.  We conclude with a summary and some speculations on a possible 4D interpretation  of the ${\cal N}=2$ 3D superstring.


\section{3D Preliminaries}
\setcounter{equation}{0}

In cartesian coordinates $\left\{\bX^\mu; \mu=0,1,2\right\}$, we define the Minkowski metric $\eta_{\mu\nu}$ and alternating pseudo-tensor 
$\varepsilon^{\mu\nu\rho}$ such that 
\be
\eta = {\rm diag} \left(-1,1,1\right) \, , \qquad \varepsilon^{012} =1 \, . 
\ee
The ``light-cone components''  are
\be
X^\pm =  \frac{1}{\sqrt{2}} \left(\bX^1 \pm \bX^0 \right) \, , \qquad X= \bX^2\, . 
\ee
Similarly, the light-cone components of an arbitrary 3-vector $\bU$ are 
\be
U_\pm = \frac{1}{\sqrt{2}} \left(\bU_1 \pm \bU_0 \right) \, , \qquad U = \bU_2\, . 
\ee
We also have
\be
U^\pm = U_\mp =  \frac{1}{\sqrt{2}} \left(\bU_1 \mp \bU_0 \right) =  \frac{1}{\sqrt{2}} \left(\bU^1 \pm \bU^0 \right)\, . 
\ee
Note that 
\be
- \bU_0^2 + \bU_1^2 + \bU_2^2 \equiv \bU^2 = 2U_+U_-  + U^2\, . 
\ee
We will make use of  the following 3D vector algebra relations for  arbitrary 3-vectors $\bU$ and $\bV$: 
\be
\bU \cdot \bV = \bU^\mu \bV^\nu \eta_{\mu\nu} \, , \qquad 
\left[\bU \wedge \bV\right]^\mu = \varepsilon^{\mu\nu\rho} \bU_\nu\bV_\rho\, , \qquad \bU_\mu = \eta_{\mu\nu} \bU^\nu\, . 
\ee

\subsection{3D Dirac matrices and Majorana spinors}

A convenient choice for the 3D  Dirac matrices is 
\be\label{Diracrep}
\Gamma^0 = i\sigma_2\, , \qquad \Gamma^1 = \sigma_1\, , \qquad \Gamma^2 = \sigma_3\, .
\ee
Observe that 
\be
\Gamma^{\mu\nu\rho} \equiv \Gamma^{[\mu}\Gamma^\nu\Gamma^{\rho]}  = \varepsilon^{\mu\nu\rho} \, \bI\, . 
\ee
The Dirac matrices satisfy the identity
\be\label{Dirac-identity}
\left(\Gamma^\mu\right)^\alpha{}_{(\beta} \left(C\Gamma_\mu\right)_{\gamma\delta)} \equiv 0\, , 
\ee
where $C$ is the antisymmetric charge conjugation matrix satisfying $C\Gamma_\mu C^{-1} = -\Gamma_\mu^T$.

A Majorana spinor is a 2-component spinor such that
\be
\bar\psi \equiv \psi^\dagger \Gamma^0 = \psi^T C\, . 
\ee
For the above representation of the Dirac matrices we may choose
\be
 C=\Gamma^0\, ,  
\ee
in which case a Majorana spinor is a real  $Sl(2;\bR)$ doublet.  For any {\it commuting} Majorana spinor $\psi$, the identity (\ref{Dirac-identity}) implies that
\be
\Gamma^\mu \psi \left(\bar\psi\Gamma_\mu \psi\right) \equiv 0\, .
\ee

The Dirac matrices in the light-cone basis are 
\be
\Gamma^\pm = \frac{1}{\sqrt{2}} \left( \Gamma^1 \pm \Gamma^0\right) \, , \qquad \Gamma \equiv  \Gamma_2 = \sigma_3\, . 
\ee
These satisfy 
\be
\left(\Gamma^\pm\right)^2 = 0\, , \qquad \Gamma^\pm \Gamma^\mp = 1\pm \sigma_3\, . 
\ee
As for vectors,  $\Gamma^\pm = \Gamma_\mp$. 

\subsection{Poincar\'e and super-Poincar\'e invariants}

The 3D Poincar\'e  group is generated by  the 3-momentum ${\cal P}_\mu$ and Lorentz 3-vector ${\cal J}^\mu$  with  non-zero commutators
\be
\left[{\cal J}^\mu, {\cal J}^\nu\right] = i \varepsilon^{\mu\nu\rho} J_\rho\, , \qquad 
\left[{\cal J}^\mu, {\cal P}^\nu\right] = i \varepsilon^{\mu\nu\rho} P_\rho\, .
\ee
In the  light-cone basis this becomes
\begin{eqnarray}\label{Poinc}
\left[{\cal J}^+,{\cal J}^- \right] &=& i{\cal J}\, , \qquad \left[{\cal J}, {\cal J}^\pm\right] = \pm i {\cal J}^\pm\, , \nonumber\\
\left[{\cal J}^\pm, {\cal P}_\mp \right] &=& \pm i{\cal P}\, , \qquad \left[{\cal J}, {\cal P}_\pm\right] = \mp i {\cal P}_\pm\, . 
\end{eqnarray}
There are two Poincar\'e Casimirs:
\be
M^2 \equiv - {\cal P}^2  \, , \qquad   \Lambda =  {\cal P}_\mu{\cal J}^\mu \, . 
\ee
Unitary irreps of the Poincar\'e group are labelled by the values of these Casimirs \cite{Binegar:1981gv}. In principle, $M^2$ may be negative but  only irreps with $M^2\ge0$ are 
physical. 
We may therefore assume that $M$ is real and non-negative. When $M>0$ we 
define the ``relativistic helicity'', which we usually abbreviate to ``helicity', by 
\be
s = \Lambda/M\, . 
\ee
This may take either sign, and parity flips the sign of $s$. We define $|s|$ to be the spin.  If the Lorentz group is $SO(1,2)$ then $s$ is an integer. If the Lorentz group is $Sl(2;\bR)$, which is the double cover of $SO(1,2)$, then $s$ is an integer or half-integer. If the Lorentz group is the universal cover of  $SO(1,2)$ then $s$ can be any real number.  

The ${\cal N}$-extended super-Poincar\'e algebra includes ${\cal N}$ Majorana spinor generators ${\cal Q}^\alpha_a$ ($\alpha=1,2$) with the following commutation relations with the Poincar\'e generators: 
\be
\left[{\cal P}, {\cal Q}^\alpha_a\right] =0\, , \qquad 
\left[{\cal J}^\mu,{\cal Q}^\alpha_a\right]  = - \frac{i}{2} \left(\Gamma^\mu\right)^\alpha {}_\beta  {\cal Q}^\beta_a\, . 
\ee
In addition they obey the  following anticommutation relation
\be\label{Qanticom}
\left\{{\cal Q}^\alpha_a,{\cal Q}^\beta_b\right\} = \delta_{ab}\left(\Gamma^\mu C\right)^{\alpha\beta} {\cal P}_\mu \, . 
\ee
The super-Poincar\'e Casimirs are (summation over $a=1,\dots ,{\cal N}$)
\be\label{superCasimirs}
M^2 \equiv - {\cal P}^2 \, , \qquad  \Omega \equiv {\cal P} \cdot {\cal J} + \frac{i}{4} \bar{\cal Q}_a {\cal Q}_a\, . 
\ee
We shall call 
\be
\bar s = \Omega/M
\ee
the  ``(relativistic) superhelicity'' of an ${\cal N}$-extended supermultiplet, and $|\bar s|$ its superspin.

\subsection{3D Superspace and superforms}\label{superspace}

The extension of Minkowski spacetime to ${\cal N}$-extended superspace involves the introduction of ${\cal N}$ {\it anticommuting} Majorana spinor coordinates $\left\{\Theta_a; a=1,\dots, {\cal N}\right\}$. The supersymmetry transformations are
\be
\delta_\epsilon {\bX}^\mu = i \bar\Theta_a \Gamma^\mu \epsilon_a \, ,  \qquad \delta_\epsilon \Theta_a = \epsilon_a\, , 
\ee
where $\epsilon_a$ are constant real anticommuting spinor parameters, and a sum over the index $a$ is implicit.  The factor of $i$ in the expression for $\delta_\epsilon \bX$ is needed because we use the standard convention that the complex conjugate of a product of anticommuting factors reverses the order, which gives a minus sign for the complex conjugation of a fermion bilinear if the order is not changed. 

A basis for the left-invariant differential 1-forms on superspace is provided by  $d\Theta_a$ and 
\be\label{left-inv}
\Pi^\mu = d\bX^\mu + i\bar\Theta_a \Gamma^\mu d\Theta_a\, . 
\ee
Allowing for non-constant $\epsilon$, one has
\be
\delta_\epsilon \Pi^\mu = -2id\bar\epsilon_a \Gamma^\mu \Theta_a\, , 
\ee
which confirms the invariance for constant parameters $\epsilon_a$.  

The WZ terms for the superstring can be constructed as follows \cite{Henneaux:1984mh}.  Consider, for ${\cal N}=1$, the 
following super-Poincar\'e invariant 3-form (the exterior product of forms is implicit):
\be\label{hthree1}
h_3^{{\cal N}=1} = \Pi^\mu \left(d\bar\Theta \Gamma_\mu d\Theta\right) \, . 
\ee
The identity (\ref{Dirac-identity}) implies that this 2-form is closed.  It is also exact,  in de Rham cohomology, because
\be\label{h2N1}
h_3^{{\cal N}=1} = dh_2^{{\cal N}=1} \, , \qquad h_2^{{\cal N}=1} =  -d{\bX}^\mu  \left(\bar\Theta \Gamma_\mu d\Theta\right) \, . 
\ee
However, $h_2$ is not super-Poincar\'e invariant,  and cannot be made so by the addition of any  exact 2-form,  so $h_3$ is non-trivial  
in Lie-superalgebra  (Chevally-Eilenberg) cohomology (see e.g \cite{deAzcarraga:1995jw}).  Because $h_3$ is super-Poincar\'e invariant  the super-Poincar\'e variation of $h_2$ is a closed 2-form, and this is sufficient for invariance of the  integral of $h_2$ over a string worldsheet. In fact, using the identity
\be
2\Gamma^\mu  d\Theta \ {\bar \Theta}\Gamma_\mu d \Theta \equiv \Gamma^\mu \Theta  \ d{\bar \Theta}\Gamma_\mu d \Theta\, , 
\ee
which is a consequence of  (\ref{Dirac-identity}), one finds that 
\be
\delta_\epsilon h_2^{{\cal N}=1} = d \left[ \bar\epsilon \Gamma_\mu \Theta \left(d\bX^\mu + \frac{i}{3} \bar\Theta \Gamma^\mu\Theta\right) \right] - 2 d\bar\epsilon\, \Gamma_\mu \Theta \left(d\bX^\mu - \frac{i}{3} \Theta\Gamma^\mu d\Theta\right)\, . 
\ee
This is non-zero even when $d\epsilon=0$, but  it is  then an exact  2-form. 

There is a generalization to ${\cal N}=2$ with 
\be\label{hthree}
h_3^{{\cal N}=2}  = \Pi^\mu \left(d\bar\Theta_1 \Gamma_\mu d\Theta_1 - d\bar\Theta_2 \Gamma_\mu d\Theta_2\right) \, . 
\ee
The relative minus sign is required for closure of $h_3$, which can be written as $dh_2$ with 
\be\label{htwo}
h_2^{{\cal N}=2}  = -\left(d{\bX}^\mu  + \frac {i}{2}\bar\Theta_a \Gamma^\mu d\Theta_a\right) \left(\bar\Theta_1 \Gamma_\mu d\Theta_1 - \bar\Theta_2 \Gamma_\mu d\Theta_2\right) \, . 
\ee
This 2-form is manifestly  Poincar\'e invariant but its supersymmetry variation (allowing for non-constant parameters) is
\bea
\delta_\epsilon h_2^{{\cal N}=2} &=&
 d \left[ \bar\epsilon_1\, \Gamma_\mu \Theta_1 \left(dX^\mu + \frac{i}{3} \bar\Theta_1 \Gamma^\mu d\Theta_1\right) - \bar\epsilon_2\, \Gamma_\mu \Theta_2 \left(dX^\mu + \frac{i}{3} \bar\Theta_2 \Gamma^\mu d\Theta_2\right)\right]  \nonumber \\
&&-\, 2d\bar\epsilon_1\, \Gamma_\mu \Theta_1 \left(dX^\mu -\frac{i}{3}\bar\Theta_1 \Gamma^\mu d\Theta_1 +i\bar\Theta_2\Gamma^\mu d\Theta_2\right)   \nonumber \\
&&+\, 2d\bar\epsilon_2\, \Gamma_\mu \Theta_2 \left(dX^\mu -\frac{i}{3}\bar\Theta_2 \Gamma^\mu d\Theta_2 +i\bar\Theta_1\Gamma^\mu d\Theta_1\right)\, .  
\eea
This is an exact 2-form for constant $\epsilon_a$.

\subsection{Parity}\label{subsec:parity}

Parity is a $\bZ_2$ transformation $\varPi$ that we may choose to have the following action on the coordinates of ${\cal N}=1$ superspace
\be\label{paritytrans1}
\varPi : \ \bX_2 \to -\bX_2\, , \qquad \Theta \to \Gamma_2 \Theta\, , 
\ee
with all other coordinates being inert.   For the extension to ${\cal N}=2$ (we will not need to consider ${\cal N}>2$)  we choose to define parity as the $\bZ_2$ transformation
\be\label{paritytrans2}
\varPi : \ \bX_2 \to -\bX_2\, , \qquad \Theta_1 \to \Gamma_2 \Theta_1\, , \qquad \Theta_2 \to - \Gamma_2\Theta_2\, , 
\ee
with all other coordinates being inert; these transformations imply the invariance of the 3-form $h_3$, and hence of the superstring WZ term.  The ${\cal N}=2$ superstring model to be considered here is  additionally invariant under the transformations $\Theta_a\to -\Theta_a$, separately for $a=1,2$, so we could choose to define  parity without the relative sign for the $\Theta_1$ and $\Theta_2$ transformation. However, the relative minus sign is required for 
standard parity assignments within supermultiplets  relevant to the ${\cal N}=2$ superstring spectrum, and for parity invariance of the massive ${\cal N}=2$ superparticle with a 
central charge  \cite{Mezincescu:2010gb}. 

Parity acts as the following outer automorphism of the ${\cal N}=1$ super-Poincar\'e algebra: 
\be
\varPi: \quad {\cal P}_2 \to - {\cal P}_2 \, , \quad {\cal J}^\pm \to -{\cal J}^\pm\, , \quad 
{\cal Q}_1 \to - \Gamma_2 {\cal Q}_1\, , 
\ee
with all other basis generators being inert. Similarly for ${\cal N}=2$,  but with the relative sign difference discussed above:
\be
\varPi: \quad {\cal P}_2 \to - {\cal P}_2 \, , \quad {\cal J}^\pm \to -{\cal J}^\pm\, , \quad 
{\cal Q}_1 \to - \Gamma_2 {\cal Q}_1\, ,  \quad  {\cal Q}_2 \to  \Gamma_2 {\cal Q}_2\, . 
\ee
In both cases, it follows that
\be
\varPi: \quad \Lambda \to -\Lambda\, , \qquad \Omega \to -\Omega\, . 
\ee


\subsection{3D Supermultiplets}\label{subsec:smultiplets}

In any hermitian operator realization of the super-Poincar\'e generators with  non-vanishing ${\cal P}_-$, and positive $M^2$, we may define  the new non-hermitian supercharges
\be
{\cal S}_a = \frac{1}{\sqrt{\sqrt{2}{\cal P}_-}} \left[ \sqrt{2} {\cal P}_- {\cal Q}_a^1 - \left({\cal P}_2 -iM\right){\cal Q}_a^2\right]\, . 
\ee
These  have the remarkably simple anticommutation relations
\be\label{simple}
\left\{ {\cal S}_a, {\cal S}_b\right\} =0\, , \qquad \left\{ {\cal S}_a, {\cal S}_b^\dagger\right\} = 2\delta_{ab} \, M^2\, . 
\ee
They also have simple commutation relations with the Poincar\'e invariant $\Lambda$
\be
\left[\Lambda, {\cal S}\right] = -\frac{1}{2} M {\cal S}\, , \qquad 
\left[\Lambda, {\cal S}^\dagger \right] = \frac{1}{2} M {\cal S}^\dagger\, , 
\ee
which shows  that the action of any of ${\cal S}_a$ on a helicity eigenstate lowers the helicity by $1/2$, whereas the action of any of ${\cal S}_a^\dagger$ raises it by $1/2$. Of course, ${\cal S}_a$ commutes with the super-invariant $\Omega$, which can be written for $M\ne0$ as 
\be
\Omega = \Lambda + \frac{1}{8M} \sum_a \left[{\cal S}_a,{\cal S}_a^\dagger\right]\, . 
\ee
It follows immediately from this formula that the value of $\Omega$ for a given supermultiplet is the
average of the values of $\Lambda$, and hence that $\bar s$ is the average of the helicities $s$. 

Irreducible supermultiplets are built by the action of the operators ${\cal S}_a^\dagger$ on a ``Clifford vacuum'' state $|\rangle$ that is annihilated  by the ${\cal S}_a$:
\be
{\cal S}_a |\rangle =0 \qquad (a= 1,\dots,{\cal N})\, . 
\ee
This gives a supermultiplet of states
\be
\left(|\rangle \, , \quad {\cal S}_a^\dagger |\rangle \, , \quad {\cal S}_a^\dagger{\cal S}_b^\dagger |\rangle\, , \dots \, , 
{\cal S}_1^\dagger \cdots {\cal S}_{\cal N}^\dagger |\rangle\right)\, . 
\ee
If the first of these states has relativistic helicity $h$ then we get a supermultiplet of $2^{\cal N}$ states with helicities ranging from $h$ to $h+ {\cal N}/2$, and `binomial' multiplicities.  As the superhelicity  is the average of the helicities,  the $\bar s=h$ supermultiplet is  the  $\bar s =0$ supermultiplet with all  component helicities shifted by $h$. 
For example, for ${\cal N}=1$, the $\bar s=h$ supermultiplet  has helicities
\be
s= \left(h-\frac{1}{4}, h+ \frac{1}{4}\right)\, . 
\ee
This is an anyon supermultiplet when $h - \frac{1}{4} \notin \bZ$. The special case of $2h \in \bZ$ yields semion supermultiplets; first studied for  $h=\frac{1}{2}$ in \cite{Sorokin:1992sy}. The $h=0$ case yields the spin-$\frac{1}{4}$ supermultiplet with $s= \left(-\frac{1}{4}, \frac{1}{4}\right)$; this has arisen  in a number of distinct contexts \cite{Witten:1999ds,Pedder:2008je,Mezincescu:2010yq}, including the level-$2$ spectrum of the ${\cal N}=1$ 3D string \cite{Mezincescu:2010yp},  because it is the unique parity-invariant   irreducible ${\cal N}=1$ supermultiplet.  The generic anyon supermultiplet has been studied in  \cite{Gorbunov:1997ie}.

 For  ${\cal N}=2$, the  $\bar s=h$ supermultiplet  has helicities
\be
s= \left(h-\frac{1}{2}, h,h, h+\frac{1}{2}\right)\, . 
\ee
Again, the superhelicity is the average of the helicities in the supermultiplet.  For $h=0$ we get the parity-invariant 3D scalar supermultiplet with spin-$0$ and spin-$1/2$ states. {\it In the absence of a central charge}, this is the unique parity-invariant 
${\cal N}=2$ supermultiplet.

When $M=0$ the ${\cal N}$ charges ${\cal S}_a$ are hermitian. These mutually-anticommuting hermitian charges also anticommute with the remaining  ${\cal N}$ linearly independent hermitian supercharges, $Q_a$, which we may choose such that 
$\left\{Q_a,Q_b\right\} = 2\delta_{ab}$.  The charges ${\cal S}_a$ annihilate the states of an  irreducible representation  of the 
super-Poincar\'e group, which are acted upon non-trivially only by the $Q_a$. For ${\cal N}=1$, there is only one charge $Q$, 
satisfying $Q^2=1$. In this exceptional case there is a trivial realization of $Q$ by the identity, but if there exists an operator $(-1)^F$ that anticommutes with $Q$ (as is the case for the ${\cal N}=1$ superstring considered here)  then the minimal realization is 2-dimensional: one bose state and one fermi state \cite{Mezincescu:2010yq}.  Although spin is not defined for massless particles, there are still two distinct unitary irreps of the Poincar\'e group corresponding to the distinction between bosons and fermions \cite{Binegar:1981gv,Deser:1991mw}.   For ${\cal N}=2$ there are two charges $Q_1$ and $Q_2$ that are realized non-trivially  and the minimal realization is again 2-dimensional but if there exists an operator $(-1)^F$ that anticommutes with $Q_1$ and $Q_2$ (as is the case for the ${\cal N}=2$ superstring considered here) then the 2-dimensional realization is complex so there are two boson and two fermion states, which is also what one finds from quantization of the ${\cal N}=2$  massless 3D superparticle \cite{Mezincescu:2010yq}.

\subsubsection{Central charges}

The ${\cal N}$-extended super-Poincar\'e algebra admits central charges for ${\cal N}\ge2$.  For the ${\cal N}=2$ case, which is of potential relevance in light of comments that we make in the conclusions,  the anticommutator (\ref{Qanticom}) becomes
\be\label{QanticomCC}
\left\{{\cal Q}^\alpha_a,{\cal Q}^\beta_b\right\} = \delta_{ab}\left(\Gamma^\mu C\right)^{\alpha\beta} {\cal P}_\mu + \varepsilon_{ab} C^{\alpha\beta} Z\, , 
\ee
where $Z$ is a real  central charge. This modification implies that (\ref{simple}) is modified to 
\be\label{simpleCC}
\left\{ {\cal S}_a, {\cal S}_b\right\} =0\, , \qquad \left\{ {\cal S}_a, {\cal S}_b^\dagger\right\} = 2M \left(\delta_{ab}\, M -i\varepsilon_{ab}\,  Z\right) 
\, .
\ee
Unitarity requires that
\be
M\ge |Z|\, . 
\ee
An ${\cal N}=2$ massive parity-preserving superparticle model in which this bound is saturated was presented in \cite{Mezincescu:2010gb}.  In the quantum theory this describes a centrally-charged parity-invariant semion supermultiplet with helicity states
$s= (-\frac{1}{4},-\frac{1}{4},\frac{1}{4},\frac{1}{4})$. In the $Z\to0$ limit, both $M$ and  $\Lambda$ go to zero, and  the helicity  
$\Lambda/M$ becomes ill-defined; the 4 massive states become the two massless bosonic and two massless fermionic states of a massless ${\cal N}=2$ supermultiplet. 

\subsection{The 3D massive particle}

The Hamiltonian form of the time-reparametrization invariant action for a point particle of non-zero mass $m$ and 
relativistic helicity $s$ is
\be\label{hamform}
S[\bX,\bP] = \int d\tau \left\{ \dot{\bX}^\mu \bP_\mu - \frac{1}{2}\ell\left(\bP^2  + m^2\right)\right\} -  sS_{LWZ}
\ee
where $\bP^2 = \eta_{\mu\nu}\bP^\mu\bP^\nu$ and $S_{LWZ}$ is  the (parity violating)  ``Lorentz Wess-Zumino'' (LWZ)
term constructed from the 
Poincar\'e-invariant closed 2-form \cite{Schonfeld:1980kb}
\begin{equation}
\frac{1}{2}\left(-{\bP}^2\right)^{-\tfrac{3}{2}} \varepsilon^{\mu\nu\rho}\,  \bP_\mu d\bP_\nu d\bP_\rho\, . 
\end{equation}
By construction, the action is Poincar\'e invariant. The Noether charges are
\be\label{poinc}
\mathcal{P}_\mu = \bP_\mu \, , \qquad \mathcal{J}^\mu =\left[{\bX}\wedge {\bP}\right]^\mu - \frac{s}{m} \bP^\mu\, .  
\ee

The time reparametrization invariance is equivalent to gauge  invariance under the  infinitesimal ``$\alpha$-symmetry'' transformation  
\be\label{alphasym}
\delta_\alpha {\bX}^\mu = \alpha{\bP}_\mu\, , \qquad \delta_\alpha {\bP}_\mu =0\, , \qquad 
\delta_\alpha \ell = \dot \alpha\, ,  
\ee
with arbitrary parameter $\alpha(\tau)$. To quantize, we must deal with this gauge invariance. As our purpose here is to illustrate some features of the light-cone gauge fixing that we will use for strings, we proceed in this way by setting
\be
X^+ = \tau \, . 
\ee
This fixes the $\alpha$-gauge invariance of (\ref{alphasym}) provided that $P_-\ne0$; which is the case for any solution
of the equations of motion as long as $m\ne0$. We may then solve the mass-shell constraint for $P_+$, which is minus the Hamiltonian in the chosen gauge:
\be
H= -P_+  = \frac{1}{2P_-} \left(P^2+m^2\right)\, . 
\ee
The light-cone gauge action naturally depends on $s$ but the $s$-dependence can be removed (following the procedure of \cite{Mezincescu:2010yq}) by defining the new variable
\be
Y^- = X^- - \frac{\Lambda P}{m^2P_-}\, , \qquad \Lambda= sm\,. 
\ee
The light-cone gauge action then becomes
\be
S[X,X^-;P,P_-] = \int d\tau \left\{ \dot X P + \dot Y^- P_-  - H \right\} \, . 
\ee
The Poincar\'e charges  (\ref{poinc})  in the light-cone gauge are 
\begin{eqnarray}\label{lcg-poinc}
\mathcal{P} &=& P\, , \qquad \mathcal{P}_- = P_- \, , \qquad \mathcal{P}_+ = -H \, , \nonumber \\
\mathcal{J} &=& Y^- P_- + \tau H\, , \qquad \mathcal{J}^+ = \tau P - X P_-\, , \nonumber \\
\mathcal{J}^- &=&  -Y^- P - XH + \Lambda/P_- \, . 
\end{eqnarray}
The $s$-dependence is now entirely in ${\cal J}^-$ and it is easily checked that ${\cal P}_\mu {\cal J}^\mu = \Lambda$, confirming that the particle has helicity $s$.  The equations of motion imply that the Poincar\'e   charges are time-independent; the explicit time-dependence is canceled by the implicit time-dependence due to the equations of motion.  

Upon quantization we have the equal-time commutation relations (we set $\hbar=1$)
\be\label{etcr}
[Y^-,P_-] = i\, , \qquad [X,P] =i\, . 
\ee
There are now operator ordering ambiguities in the expressions for $\mathcal{J}$ and $\mathcal{J}^-$. These ambiguities are fixed by the twin requirements of  hermiticity and   closure of the Lorentz algebra. The quantum Lorentz generators are
\begin{eqnarray}
\mathcal{J} &=& \frac{1}{2}\left\{Y^-, P_-\right\} + \tau H \, , \qquad \mathcal{J}^+ = \tau P -Y P_-\, , \nonumber \\
 \mathcal{J}^- &=&  -Y^- P - \frac{1}{2}\left\{X,H\right\} + \Lambda/P_-\, . 
\end{eqnarray}
It should now be understood that the canonical variables in these expressions are operators, as is $H$.   Again, the explicit time-dependence is cancelled by the implicit time-dependence of the operators.  Using the equal-time commutation relations (\ref{etcr})
one may verify that the commutation relations (\ref{Poinc}) are satisfied, and hence that the quantum theory preserves the Poincar\'e invariance of the classical theory. This was to be expected but  virtually the same computation is what is needed to verify Poincar\'e invariance for the 3D string. The only difference is in the form of the Hamiltonian $H$ and the Poincar\'e invariant $\Lambda$. As long as these operators commute, one finds that the commutation relations (\ref{Poinc}) are obeyed,
so the proof of Lorentz invariance for the 3D string will reduce to checking that $[H,\Lambda]=0$.


\section{The 3D bosonic string}
\setcounter{equation}{0}

The Nambu-Goto action for the closed bosonic 3D string of tension $T$ is
\be\label{Nambu-Goto}
S[\bX] = -T\!\int \! d\tau \oint \frac{d\sigma}{2\pi} \sqrt{\left(\dot{\bX} \cdot {\bX}^\prime\right)^2 - \dot\bX^2 \left({\bX}^\prime\right)^2}\, , 
\ee
where an  overdot indicates a derivative with respect to the arbitrary time parameter $\tau$ and a prime indicates a derivative with respect to the arbitrary string coordinate $\sigma$, which we assume to  be identified with  $\sigma + 2\pi$. This action involves the background 3D Minkowski metric $\eta$ through the scalar product.  

Our starting point, however, is not the Nambu-Goto action but rather the following Hamiltonian version, with 
3-momentum $\bP$  and Lagrange multipliers $\ell$ (for the time-reparametrization constraint) and $u$ (for the $S^1$-diffeomorphism constraint):
 \be\label{closedstring}
S[\bX,\bP;\ell,u] = \int d\tau \oint \frac{d\sigma}{2\pi} \left\{\dot {\bX}^\mu {\bP}_\mu     - \frac{1}{2}\ell \left[{\bP}^2 + (T{\bX}^\prime)^2\right]     
 - u\,  {\bX}^{\prime \mu} {\bP}_\mu  \right\}\, . 
\ee
Although the Lagrange multipliers should be viewed here as unrestricted variables, it is necessary to assume that $\ell$ is nowhere zero in order to recover the Nambu-Goto action (\ref{Nambu-Goto}), by progressive elimination of $\bP$, $\ell$ and $u$. For this reason, the Hamiltonian formulation is not strictly equivalent to the Nambu-Goto formulation, even classically, but it should be appreciated that the classical action is merely a starting point for the construction of a quantum theory,  and the Hamiltonian formulation of the Nambu-Goto string is convenient for this purpose.

The action (\ref{closedstring}) is invariant under the Poincar\'e transformations generated by the Noether charges
\be\label{noether1}
{\cal P}_\mu = \oint\! \frac{d\sigma}{2\pi} \,  {\bP}_\mu \, , \qquad 
{\cal J}^\mu = \oint\! \frac{d\sigma}{2\pi}  \left[ {\bX}\wedge {\bP}\right]^\mu\, . 
\ee
It is also invariant under the discrete parity transformation
\be\label{parity1}
\bX_2 \to - \bX_2 \, , \qquad \bP_2\to -\bP_2\, , 
\ee
with all other variables being inert.  In addition to these rigid invariances, the action (\ref{closedstring}) is  invariant under the following infinitesimal  gauge transformations with parameters $\alpha,\beta$:
\begin{eqnarray}\label{alphabetasym}
\delta \bX &=&  \alpha \bP + \beta {\bX}^\prime \, , \nonumber \\
\delta {\bP} &=& T^2(\alpha{\bX}^\prime)^\prime + \left(\beta {\bP}\right)^\prime\, , \nonumber \\
\delta \ell &=& \dot\alpha + u^\prime \alpha -u \alpha^\prime + \left(\ell^\prime \beta - \ell\beta^\prime\right) \nonumber\\
\delta u &=& \dot\beta + u^\prime \beta - u\beta^\prime + T^2\left(\alpha \ell^\prime - \ell\alpha^\prime\right) \, . 
\end{eqnarray}
Note that not only is the action gauge invariant but so also are the Noether charges.


\subsection{Light-cone gauge } 

We now introduce the light-cone coordinates $(X^+,X^-,X)$ and their conjugate momenta $(P_+,P_-,P)$. The light-cone gauge is defined by the choice
\be\label{lcg}
X^+ =\tau \, , \qquad P_- = p_-(\tau)\, , 
\ee
where $p_-(\tau)$ is a  function of $\tau$ only that we assume to be nowhere zero\footnote{For  $p_-=0$ one gets  longitudinal `kink' modes \cite{Bardeen:1975gx} which are presumably  related to  the Liouville mode that arises in the Polyakov approach  from a {\it quantum} conformal anomaly. In the light-cone gauge this anomaly is pushed into an anomaly of the Lorentz algebra,  so one might expect to have to include these modes to get a Lorentz invariant theory in a sub-critical dimension, as is the case for the 2D string \cite{Bardeen:1975gx}. However, we shall see that no longitudinal modes (whether of quantum or classical  origin) are 
needed for Lorentz invariance of the 3D string.}. 
This gauge choice leaves only the  residual global gauge invariance 
induced by a constant shift of $\sigma$: 
\be\label{residual}
\delta_{\beta_0} \bX = \beta_0\,  {\bX}^\prime\, ,  \qquad \delta_{\beta_0} u = \dot \beta_0 + u^\prime \beta_0 \, , \qquad 
\delta_{\beta_0} \ell = \beta_0 \ell^\prime\, , 
\ee
where
\be
\beta_0(\tau) =  \oint \! \frac{d\sigma}{2\pi}\, \beta \, . 
\ee
To obtain the action in light-cone gauge,  we first define
\begin{eqnarray}
x(\tau) &=& \oint \! \frac{d\sigma}{2\pi} \, X\, , \qquad  x^-(\tau) = \oint \! \frac{d\sigma}{2\pi}\,  X^-\, \nonumber \\
p(\tau) &=&  \oint \! \frac{d\sigma}{2\pi} \, P\, , \qquad p_+(\tau) =  \oint \! \frac{d\sigma}{2\pi}\,  P_+\, , 
\end{eqnarray}
and 
\begin{eqnarray}
\bar X &=& X-x\, , \qquad \bar X^- = X-x^- \, , \nonumber\\
\bar P &=& P-p \, ,  \qquad \bar P_+  = P_+ - p_+ \, , 
\end{eqnarray}
and also
\be
u_0 = \oint \! \frac{d\sigma}{2\pi}\,  u \, , \qquad \bar u = u -u_0\, . 
\ee
Using the gauge conditions (\ref{lcg}), we now find that the string Lagrangian reduces to
\begin{eqnarray}\label{halfway}
L &=& \dot x p  + \dot x^- p_-  + p_+ 
+  \oint \! \frac{d\sigma}{2\pi}\,   \dot {\bar X} \bar P   -  u_0 \oint \! \frac{d\sigma}{2\pi} \, {\bar X}^\prime P 
- \oint \! \frac{d\sigma}{2\pi}\,  \bar u \, \bar X' P \nonumber \\
&& + \ p_-\oint \! \frac{d\sigma}{2\pi}  \left\{ \bar X^- \bar u^\prime -  \ell\left(P_+ + \frac{1}{2p_-} 
\left[ P^2 + (TX^\prime)^2\right]\right) \right\}\, . 
\end{eqnarray}
In this form of the action, $\bar X^-$ is a Lagrange multiplier imposing the constraint $\bar u'=0$, which implies $\bar u=0$. The constraint imposed by the lapse function $\ell$  is also easily solved: 
\be\label{lapsecon}
P_+ = -\frac{1}{2p_-} \left[ P^2 + (TX^\prime)^2\right]\, . 
\ee 
This leads to the Lagrangian density 
\be\label{lcgLag}
L = \left\{\dot x p  + \dot x^- p_- +  \oint \! \frac{d\sigma}{2\pi} \, \dot {\bar X} \bar P \right\}  - H
- u_0\oint \! \frac{d\sigma}{2\pi}\,  \bar X' \bar P \,  , 
\ee
where the Hamiltonian is 
\be\label{Ham}
H \equiv -p_+ = \frac{1}{2p_-} \left( p^2 + {\cal M}^2\right)\, , 
\ee
with
\be
{\cal M}^2 =  \oint \! \frac{d\sigma}{2\pi}  \left[ \bar P^2 + (T{\bar X}^\prime)^2\right]\, . 
\ee

As expected, there is a residual global constraint imposed by $u_0$, corresponding to the residual global gauge invariance 
which is now just
\be\label{residual2}
\delta_{\beta_0}\phi = \beta_0 \phi^\prime \, , \qquad \delta_{\beta_0} u_0 = \dot \beta_0\, . 
\ee
The $u_0$-dependence of the Lagrangian (\ref{lcgLag}) converts derivatives with respect to $\tau$ into covariant $\tau$ derivatives, defined for any dynamical variable $\phi$ by
\be
D_\tau \phi = \dot \phi - u_0 \phi^\prime\, . 
\ee
This transforms covariantly under (\ref{residual2}):
\be
\delta_{\beta_0}\left(D_\tau \phi\right) = \beta_0 \left(D_\tau \phi\right)^\prime\, . 
\ee
Using this notation, the  light-cone-gauge action may now be written in a form that is  manifestly invariant under this 
residual gauge invariance: 
\be\label{lcgaction}
S = \int d\tau \left\{ \dot x p  + \dot x^- p_- +  \oint \! \frac{d\sigma}{2\pi} \, \bar P D_\tau \bar X   - H\right\}\, .
\ee 
This action is clearly still invariant under the parity transformation (\ref{parity1}), which now reads
\be
X\to -X\, , \qquad P\to -P\, , 
\ee
with all other variables being inert. It is also still Poincar\'e invariant, despite appearances: the infinitesimal transformations are easily found by  working out the compensating gauge transformations needed to maintain the gauge choice when performing an infinitesimal 
Poincar\'e transformation, and these transformations can then be used to find the Noether charges. However, because the Noether charges  are gauge-invariant, one finds the same result by simply substituting the gauge-fixing conditions into the expressions (\ref{noether1}). This gives
\be\label{calPs}
{\cal P}_2= p\,  , \qquad {\cal P}_- = p_- \, , \qquad {\cal P}_+ = -H\, ,
\ee
and
\be\label{LorentzString}
{\cal J} =  x^- p_- + \tau H\, , \qquad {\cal J}^+ = \tau p  -x p_-\, , \qquad 
{\cal J}^- =  -x^- p -xH  + \varLambda/p_- \, , 
\ee
where
\be\label{lam}
\varLambda = p_- \oint\! \frac{d\sigma}{2\pi} \left[ \bar X {\bar P}_+ - \bar X^- \bar P\right] \, . 
\ee
One may verify that all these charges are time-independent as a consequence of the equations of motion.  
The two Poincar\'e invariants are
\be\label{poincinv}
{\cal P}^2 =- {\cal M}^2 \, , \qquad {\cal P} \cdot{\cal J} =  \varLambda\, . 
\ee

Observe that  $\varLambda$ depends on $\bar X^-$ as well as the canonical variables of the final action, but the  equation of motion of  
$\bar u $ in (\ref{halfway}) is
\be\label{Xminus}
p_- \left(\bar X^-\right)^\prime + p \bar X^\prime = -\bar X^\prime \bar P + \oint \frac{d\sigma}{2\pi} \bar X^\prime \bar P \, , 
\ee
which  will allow us to express  $\bar X^-$ in terms of  $(p_-,p)$ and the Fourier coefficients of $(\bar X, \bar P)$.


\subsection{Fourier expansion} 

We see from (\ref{lcgLag}) that the physical variables in the light-cone gauge are the canonical pairs $(x,p)$, $(x^-,p_-)$ and either  $(\bar X,\bar P)$ or the coefficients in their Fourier expansions. As is standard, we actually choose to Fourier expand the combinations 
$\bar P \pm T\bar X^\prime$:
\begin{eqnarray}\label{separate}
\bar P -T\bar X^\prime &=& \sqrt{2T} \sum_{n=1}^\infty \left[e^{in\sigma} \alpha_n + e^{-in\sigma}  \alpha_n^*\right] \, , \nonumber \\
\bar P +T\bar X^\prime &=& \sqrt{2T} \sum_{n=1}^\infty \left[e^{in\sigma} {\tilde \alpha}_n^* + e^{-in\sigma} {\tilde \alpha}_n\right]\, . 
\end{eqnarray}
This implies that 
\begin{eqnarray}
\bar X &=& \frac{i} {\sqrt{2T}} \sum_{n=1}^\infty \frac{1}{n}\left[ e^{in\sigma} \left(\alpha_n - \tilde \alpha_n^*\right) - e^{-in\sigma}\left(\alpha_n^* -\tilde \alpha_n\right)\right] \, , 
\nonumber \\
\bar P &=&  \sqrt{\frac{T}{2}}\sum_{n=1}^\infty\left[ e^{in\sigma} \left(\alpha_n +\tilde \alpha_n^*\right) + e^{-in\sigma}\left(\alpha_n^* + \tilde \alpha_n\right)\right]\, . 
\end{eqnarray}
It follows from the first of these expressions that 
\be
\bar X' = -\sqrt{\frac{1}{2T}}  \sum_{n=1}^\infty \left[ e^{in\sigma} \left(\alpha_n - \tilde \alpha_n^*\right) + e^{-in\sigma} \left(\alpha_n^* -\tilde \alpha_n\right)\right]\, , 
\ee
and hence that the Lagrangian (\ref{lcgLag}) may be written as
\be\label{oscLag}
L= \left\{\dot x p  + \dot x^- p_- + i \sum_{n=1}^\infty \frac {1}{n}\left[ \dot \alpha_n \alpha_n^* + \dot{\tilde \alpha}_n \tilde \alpha_n^* \right] \right\} -H + u_0 \sum_{n=1}^\infty \left[\alpha_n^* \alpha_n - \tilde \alpha_n^*\tilde \alpha_n\right]\, , 
\ee
where the Hamiltonian is as in (\ref{Ham}) but now ${\cal M}^2$ is expressed as a sum over Fourier modes
\be
{\cal M}^2 =  2T \sum_{n=1}^\infty \left[ \alpha_n^* \alpha_n + \tilde \alpha_n^*\tilde \alpha_n\right] \, .
\ee
Note that parity now acts as
\be
x \to -x\, , \qquad p\to -p\, , \qquad \alpha_n \to -\alpha_n \, , \qquad \tilde\alpha_n \to - \tilde\alpha_n\, .
\ee

To obtain expressions for the Lorentz generators (\ref{LorentzString}) in terms of the same variables, we need an expression in terms of them  for  $\bar X^-$. To this end we  use (\ref{Xminus}) to deduce that
\be\label{XminusExpansion}
\bar X^-  = - \frac{1}{p_-} \left\{ p\bar X + \sum_{n=1} ^\infty \frac{i}{n}\left[ e^{in\sigma} 
\left(\beta_n - \tilde \beta_n^*\right) - e^{-in\sigma}\left(\beta_n^* -\tilde \beta_n\right)\right] \right\}\, , 
\ee
where
\be\label{bn}
 \beta_n = \frac{1}{2} \sum_{m=1}^{n-1} \  \alpha_m \alpha _{n-m}+ \sum_{m>n}  \alpha _m\alpha_{m-n}^*\, , 
\ee
and similarly for $\tilde \beta_n$.   The $\beta_n$ and $\tilde\beta_n$ coefficients also arise in the Fourier expansion of 
$\bar P_+$,  as given in  (\ref{lapsecon}):
\be\label{Pbarplus}
\bar P_+  = - \frac{1}{p_-} \left\{p \bar P +   T\sum_{n=1}^\infty  \left[ e^{in\sigma} 
\left(\beta_n + \tilde \beta_n^*\right) + e^{-in\sigma} \left(\beta_n^* + \tilde \beta_n\right) \right] \right\}\, . 
\ee

We now have Fourier expansions for each of the variables appearing in the integrand of the expression (\ref{lam}) for $\varLambda$. Using them, we deduce that 
\be
\varLambda = \varLambda_+ \, + \, \varLambda_- \, , \qquad \varLambda_+ = \sqrt{2T}\, \lambda\, , \qquad 
\varLambda_- = \sqrt{2T}\, \tilde\lambda\, , 
\ee
where
\be\label{littlelambdas}
\lambda =  \sum_{n=1}^\infty \frac{i}{n} \left(\alpha_n^* \beta_n - \beta_n^*\alpha_n  \right) \, , \qquad
\tilde\lambda =  \sum_{n=1}^\infty \frac{i}{n} \left(\tilde\alpha_n^* \tilde\beta_n - \tilde\beta_n^*\tilde\alpha_n  \right)\, . 
\ee


\subsubsection{Equations of motion}

Before moving on to the quantum theory, we comment on the equations of motion. The Lagrangian  (\ref{oscLag}) leads to the equations of motion
\be
\dot p=\dot p_- =0 \, , \qquad \dot x = p/p_- \, , \qquad \dot x^- =-H/p_-\, , 
\ee
and 
\be
D_\tau \alpha_n = -i n\omega \alpha_n\, , \qquad 
D_\tau \tilde\alpha_n = -i n\omega  \tilde\alpha_n\, , 
\ee
where
\be
\omega = T/p_-\, . 
\ee
Using these equations, and the expression (\ref{Pbarplus}) for $P_+$, one can show that the expression (\ref{XminusExpansion}) for $\bar X^-$ implies that
\be\label{dotXminus}
p_- D_\tau \bar X^- = \bar P_+\, . 
\ee

In the gauge $u_0=0$,  the equations for $(\alpha_n,\tilde\alpha_n)$ have the solution
\be
\alpha_n(\tau) = \alpha_n(0)  e^{-in\omega \tau}\, , \qquad 
\tilde\alpha_n(\tau) = \tilde\alpha_n(0) e^{-in\omega\tau}\, , 
\ee
which gives
\begin{eqnarray} 
\bar P - T\bar X^\prime &=& \sqrt{2T} \sum_{n=1}^\infty \left[ e^{in\left[\sigma - \omega\tau\right]} \alpha_n(0)+ c.c. \right]  \nonumber \\
\bar P + T\bar X^\prime &=& \sqrt{2T} \sum_{n=1}^\infty \left[ e^{-in\left[\sigma + \omega\tau\right]}\tilde\alpha_n(0) + c.c. \right]\, . 
\end{eqnarray}

This confirms that the $\alpha_n$ are the Fourier coefficients for right-moving modes and $\tilde\alpha_n$ the Fourier coefficients for left-moving modes, but one might have expected to find that  $\omega=1$ since waves on the string  travel along it  at the speed of light (which is $c=1$ in the units used here).  However, the scale associated with the time variable $\tau$ is arbitrary, and this is reflected in the arbitrariness of the angular frequency $\omega= T/p_-$. Note that $p_-$ is set to a constant by the equations of motion. A natural choice is 
\be
p_- = T 
\ee
since this implies that $\omega=1$. However, it is important not to set $p_-=T$ in the action; doing so would cause the $\dot x^- p_-$ term to become an irrelevant total derivative and the action would no longer be Lorentz invariant. It is also important not to set 
$p_-=T$ in the expressions for the Noether charges, at least before  evaluation of Poisson brackets (classically) or commutators (quantum mechanically).

\subsection{Quantum bosonic string}

The non-zero Poisson brackets of the canonical variables in the light-cone-gauge action (\ref{lcgaction}) are
\be
\left\{ x,p\right\}_{pb} =1 \, , \qquad \left\{ x^-,p_-\right\}_{pb} =1\, , \qquad 
\left\{ X(\sigma), P(\sigma')\right\}_{pb} = 2\pi \, \delta(\sigma-\sigma')\, . 
\ee
In the quantum theory, these variables are promoted to operators with the commutation relations (we set $\hbar=1$)
\be
\left[x,p\right] = i \, , \qquad \left[x^-,p_-\right] = i \, , \qquad \left[X(\sigma), P(\sigma')\right] = 2\pi i \, \delta(\sigma-\sigma')\, . 
\ee
The last of these can be achieved by promoting to operators the Fourier coeficients $(\alpha_n,\tilde\alpha_n)$ so that the non-zero commutators are 
\be
\left[\alpha_n, \alpha_n^\dagger \right] = n \, , \qquad \left[ \tilde \alpha_n, \tilde \alpha_n^\dagger \right] = n\,  ,  \qquad  \forall\,  n \in \bZ^+\, . 
\ee
The quantum Hamiltonian is then
\be\label{msquared}
H = \frac{1}{2p_-} \left(p^2 + {\cal M}^2\right)\, , \qquad {\cal M}^2 =  2T\left(N+\tilde N -a\right)\, ,
\ee
where $a$ is an arbitrary constant arising from operator ordering ambiguities, and the ``level-number'' operators $N$ and $\tilde N$ are
\be
N= \sum_{n=1}^\infty \alpha_n^\dagger \alpha_n,  \qquad{\tilde N}= \sum_{n=1}^\infty {\tilde\alpha}_n^\dagger {\tilde \alpha}_n\, . 
\ee
The constraint imposed by $u_0$ is the level-matching  condition, which must be  imposed as a physical-state condition
in the quantum theory:  for physical state $|{\rm phys}\rangle$, 
\be\label{levelmatching}
\left(N-\tilde N\right)|{\rm phys}\rangle =0\,  .
\ee
The string ground state takes the tensor product form
\be\label{string-gs}
|p,p_-\rangle \otimes |0\rangle_+ \otimes |0\rangle_-\, , 
\ee
where $|0\rangle_+$ is the ground state for the right-moving modes and $|0\rangle_-$ is the ground state for the left-moving modes:
\be
\alpha_n |0\rangle_+ =0 \, , \qquad \tilde\alpha_n |0\rangle_- =0 \, , \qquad \forall \, n\in \bZ^+\, . 
\ee
Excited string states are found by the action of oscillator creation operators on this ground state.  Such states are eigenstates of the level operators $N$ and $\tilde N$, with eigenvalues that we also call $N$ and $\tilde N$. To be physical, these eigenstates must satisfy the level-matching condition $N=\tilde N$. We may therefore organize all physical  states according to their level $N$. In addition, 
\be\label{mass-spectrum}
{\cal M}^2|_N = 2T\left(2N-a\right) \, . 
\ee

Because of the level-matching constraint, not only is the (level-$0$) oscillator ground state unique, for given $p$ and $p_-$, but so also is the first (level-$1$)  excited state, 
\be 
\alpha_1^\dagger |0\rangle_+  \otimes \tilde \alpha_1^\dagger |0\rangle_-   \equiv   |1\rangle_+ \otimes  |1\rangle_- \, . 
\ee
There are four physical level-$2$  states, which are tensor products of 
\be\label{basis-2}
|1,1\rangle_+ = \frac{1}{\sqrt{2}} \left(\alpha_1^\dagger\right)^2 |0\rangle_+ \, , \qquad |2\rangle_+ =  \frac{1}{\sqrt{2}}\alpha_2^\dagger |0\rangle_+\, , 
\ee
with the analogous two states built on  $|0\rangle_-$. At level $3$ we need to consider the three (orthonormal basis) states
\begin{eqnarray}\label{basis-3}
|1,1,1\rangle_+  &=&  \frac{1}{\sqrt{6}}\left(\alpha_1^\dagger\right)^3 |0\rangle_+ \, , \qquad
|1,2\rangle_+ =   \frac{1}{\sqrt{2}}\alpha_1^\dagger \alpha_2^\dagger |0\rangle_+ \, ,  \nonumber \\
|3\rangle_+ &=&  \frac{1}{\sqrt{3}}\alpha_3^\dagger |0\rangle_+ \, ,
\end{eqnarray}
and this leads to a total of nine physical states. 

At level $4$ we need to consider the five (orthonormal basis) states
\begin{eqnarray}\label{basis-4}
|1,1,1, 1\rangle_+  &=&  \frac{1}{\sqrt{4!}}\left(\alpha_1^\dagger\right)^4 |0\rangle_+ \, , \qquad  \qquad  \qquad
|1,1,2\rangle_+ =   \frac{1}{{2}}\left(\alpha_1^\dagger \right)^2\alpha_2^\dagger |0\rangle_+ \, , \nonumber \\
|1,3\rangle_+ &=&  \frac{1}{\sqrt{3}}\alpha_1^\dagger\alpha_3^\dagger |0\rangle_+ \, ,   \qquad
|2,2\rangle_+  =  \frac{1}{2\sqrt{2}}\left(\alpha_1^\dagger\right)^2 \alpha_2^\dagger |0\rangle_+ \, , \nonumber  \\
|4\rangle_+ &=&   \frac{1}{{2}} \alpha_4^\dagger |0\rangle_+ \, , \qquad
\end{eqnarray}
and this leads to a total of  twenty-five level-$4$ physical states.

\subsubsection{Lorentz covariance and Parity}

As the light-cone gauge renders the classical Lorentz invariance non-manifest, there is no guarantee that the quantum string will be Lorentz invariant. We must therefore check Lorentz invariance.  The quantum translation generators are 
\be\label{quantumtrans}
{\cal P}_2= p\,  , \qquad {\cal P}_- = p_- \qquad {\cal P}_+ = -H\, ,
\ee
exactly as in (\ref{calPs}) but now with the operator Hamiltonian of (\ref{msquared}), and the quantum Lorentz generators are 
\begin{eqnarray}\label{quantumLorentz}
{\cal J} &=& \frac{1}{2}\left\{x^-, p_-\right\} + \tau H \, , \qquad {\cal J}^+ = \tau p -x p_-\, , \nonumber \\
{\cal J}^- &=&  -x^- p - \frac{1}{2}\left\{x,H\right\} + \varLambda/p_-\, .
\end{eqnarray}
Here,  $\varLambda= \varLambda_+ + \varLambda_-$, with 
\begin{eqnarray}\label{qlam}
\frac{1}{\sqrt{2T}} \varLambda_+ &=&  \lambda \equiv 
\sum_{n=1}^\infty \frac{i}{n} \left(\alpha_n^\dagger \beta_n -  \beta_n^\dagger \alpha_n \right)  \, , \nonumber \\
\frac{1}{\sqrt{2T}} \varLambda_- &=&  \tilde\lambda \equiv 
 \sum_{n=1}^\infty \frac{i}{n} \left(\tilde\alpha_n^\dagger \tilde\beta_n -   \tilde\beta_n^\dagger\tilde\alpha_n \right) \, , 
\end{eqnarray}
where $\beta_n$ and $\tilde\beta_n$ are now the operators
\begin{eqnarray}\label{betas}
\beta_n &=& \frac{1}{2} \sum_{m=1}^{n-1} \alpha_m \alpha_{n-m} + \sum_{m>n} \alpha_m \alpha_{m-n}^\dagger \, , \nonumber \\
\tilde\beta_n &=& \frac{1}{2} \sum_{m=1}^{n-1} \tilde\alpha_m \tilde\alpha_{n-m} +
 \sum_{m>n} \tilde\alpha_m \tilde\alpha_{m-n}^\dagger\, . 
\end{eqnarray}
All operator ordering ambiguities in the quantum Lorentz generators are fixed by the requirements of hermiticity and closure of the algebra. In particular, there is no Lorentz anomaly, so the quantum theory is  Lorentz invariant.  This was to be expected because
the ``dangerous commutators'' are antisymmetric in the $(D-2)$ ``transverse space''  indices and hence trivially absent for $D=3$. 
As a result, the computation is equivalent to the one that must be done for the massive particle except that one needs to check that 
$[N,\lambda]=0$, which implies $[H,\Lambda]=0$. For this step, it is convenient to first establish the commutation relations
\be\label{albet}
\left[\alpha_n,\beta_m\right] = n\alpha_{n+m}\, , \qquad \left[\alpha_n^\dagger, \beta_m\right] 
= \begin{cases} -n\alpha_{m-n} & n<m \\ 
\quad 0 & n=m \\
-n\alpha_{n-m}^\dagger & n>m 
\end{cases} \quad , 
\ee
and then to use the identity
\be
\sum_{m=1}^\infty \sum_{n=m+1}^\infty \equiv \sum_{n=2}^\infty \sum_{m=1}^{n-1}\, . 
\ee

It also remains true in the quantum theory  that 
\be
{\cal P}^2 = - {\cal M}^2\, , \qquad {\cal P}\cdot {\cal J} = \varLambda\, , 
\ee
where the operators ${\cal M}^2$ and $\varLambda$ are given by (\ref{mass-spectrum}) and (\ref{qlam}) respectively.  It is straightforward to verify that these two operators  commute with each other and with all generators of the Poincar\'e algebra. 

The parity operator of the quantum theory is\footnote{Recall that the parity operator for the harmonic oscillator is $exp(i\pi\hat N)$ where $\hat N$ is the particle number operator.}
\be\label{parityop}
\varPi = \varPi_0 \exp \left[i\pi \sum_{n=1}^\infty \frac{1}{n} \left(\alpha_n^\dagger \alpha_n + \tilde\alpha^\dagger \tilde\alpha_n\right)\right]\, , 
\ee
where
\be
\varPi_0 = \int \!dp\  |-p\rangle \langle p | \, .
\ee
The operator $\varPi$ anticommutes with all the creation and annihilation operators.  It therefore commutes with  
$N$ and $\tilde N$,  and hence with the Hamiltonian. Also, it  {\it anticommutes} with $\varLambda$. Parity is therefore 
preserved by the quantum theory, and all states of non-zero spin must appear in parity doublets of opposite-sign helicities. 
For the first few low-lying levels, this is  verified by the explicit computations to follow.

\subsubsection{Helicity spectrum} 

As ${\cal M}^2$ and $\varLambda$ commute, they are simultaneously diagonalizable. This means  that $\varLambda$ is block diagonal in a basis in which ${\cal M}^2$ is diagonal, with blocks that may be labeled by the  level number $N$. Since  $\lambda$ and $\tilde\lambda$ commute, they too may be simultaneously diagonalized.  It follows that 
\be\label{lambdasum}
\lambda = \sum_{n=2}^\infty \lambda_n \, , \qquad \tilde\lambda = \sum_{n=2}^\infty \tilde\lambda_n \, , 
\ee
where $\lambda_n$ annihilates all states with $N < n$ but not all those with $N \geq n$. The absence of $n=0$ and $n=1$ contributions to the sum is easily verified, 
and it implies that $\varLambda$ annihilates the states at $N=0$ and $N=1$; this is  expected because  these  levels each contain a single physical state which must be a parity singlet and hence a scalar.  At level $2$ we need consider only  $\lambda_2$ because $\lambda_n$ for $n\ge3$ annihilates all  states at levels $N=0,1,2$. A computation shows that
\be\label{lambda2}
\lambda_2 =  \frac{3i}{4} \left[ \left(\alpha_1^\dagger\right)^2 \alpha_2 - \alpha_2^\dagger \alpha_1^2\right] \, . 
\ee
This reduces to the matrix $(3/2)\sigma_2$ in the  level-$2$ basis  (\ref{basis-2}) so $\lambda$ has 
eigenvalues $\pm 3/2$. The same is obviously true for $\tilde\lambda$, so the eigenvalues of $\varLambda$ at level $2$ are $(0,0,3,-3)$ times $\sqrt{2T}$. We must divide by the level-$2$ mass $\sqrt{2T(4-a)}$ to get the helicities, which are  therefore
\be
s_2 =\left(0,0, \pm\frac{3}{\sqrt{4-a}}\right)\, . 
\ee
As implied by parity,  each non-zero spin  occurs twice, once for each sign of the helicity. 

At level $3$ we need $\lambda_3$ and a computation gives
\be\label{lambda3}
\lambda_3 = \frac{7i}{6} \left(\alpha_1^\dagger \alpha_2^\dagger \alpha_3 - \alpha_3^\dagger \alpha_1 \alpha_2\right)  \, .
\ee
We also need $\lambda_2$ because, for example, it does not annihilate $|1,1,1\rangle_+$, but we do not need $\lambda_4$ or higher terms. 
In the level-$3$ basis (\ref{basis-3}), one finds that $\lambda$ reduces to the matrix
\be
\frac{i}{2\sqrt{3}}\left( \begin{array}{ccc} 0 & -9 &0 \\  9 &0 & -7\sqrt{2} \\ 0& 7\sqrt{2} &0\end{array} \right)\, , 
\ee
which has eigenvalues $(0, \pm \sqrt{179/12})$. This leads to the level-$3$ helicity content
\be
s_3 = \left(0,0,0, \pm\sqrt{\frac{179}{12\left(6-a\right)}},\pm \sqrt{\frac{179}{12\left(6-a\right)}}, \pm\sqrt{\frac{179}{3\left(6-a\right)}}\right)\, . 
\ee
Observe again that non-zero helicities appear in parity doublets of opposite helicity. 

At level $4$ we need $\lambda_4$ and a computation gives
\be
\lambda_4 = {i} \left\{\frac {13}{12} \left(\alpha_3^\dagger \alpha_1^\dagger \alpha_4 - \alpha_4^\dagger \alpha_1 \alpha_3\right) + \frac {3}{8} \left[\left(\alpha_2^\dagger\right)^2 \alpha_4 - \alpha_4^\dagger \left(\alpha_2 \right)^2\right]\right\}  \, .
\ee
We also need $\lambda_3$  and $\lambda_2$. In the level-$4$ basis (\ref{basis-4}), one finds that $\lambda$ reduces to the matrix
\be
i\left(\begin{array}{ccccc} 0 & -{3\sqrt{6}/ 2} & 0 & 0 & 0 \\  
      {3\sqrt{6}/ 2}  & 0 & -{7\sqrt{3}/ 3} & -{3\sqrt{2}/2} & 0 \\
               0& {7\sqrt{3}/ 3} &0 &0 & -{13 \sqrt{3}/ 6}\\ 
                0 & 3{\sqrt{2}/ 2} &0 & 0& -{3\sqrt{2}/ 2} \\
                 0 & 0 & {13 \sqrt{3}/ 6}&{3\sqrt{2}/ 2} & 0 \end{array} \right)\, , 
\ee
which has eigenvalues 
\be
\left(0, \pm\sqrt{\frac{1}{24} \left(635+\sqrt{258505}\right)}, \pm \sqrt{\frac{1}{24} \left(635-\sqrt{258505}\right)} \right)\, . 
\ee
Or, in plain numbers,  $\{0, \pm 6.9024\cdots,\pm 2.29643\cdots \}$. The helicities are found by dividing these numbers by   $\sqrt{8 - a}$.  

We are free to choose the intercept parameter $a$ except that there are tachyons unless
$a\le0$. The choice $a=0$ is natural because this makes the ground state a massless scalar.
In this case the first excited state (level $1$) is a massive scalar, there are then  spins $(0,0,3/2)$ at level $2$ and  some irrational spin anyons  at level $3$. We shall call this the ${\cal N}=0$ string since its spectrum is of direct relevance to the spectrum of the ${\cal N}=1$ superstring. 

However,  since the ground state of the critical bosonic string is a tachyon, the $a>0$ cases
should perhaps be considered too. In particular, the choice $a=2$ leads to a tachyonic scalar ground state and a massless first-excited state, just like the critical bosonic string although the  first excited state is a scalar in 3D. For $a>2$ this scalar excited state is a tachyon too but as long as $a<4$ there are no other tachyons.  For $a>4$ there are non-scalar tachyonic  excited states in addition to the scalar ground-state tachyon.  The $a=4$ case is special because there are then states of  infinite helicity; we believe that these correspond to unitary irreps  of the 3D Poincar\'e  group that are analogs of Wigner's  unitary ``infinite spin'' (alias ``continuous spin'' ) irreps  of the 4D Poincar\'e group (see \cite{Edgren:2005gq} for a recent discussion). If so, the 3D Nambu-Goto string would provide a novel physical model for these Poincar\'e irreps; we intend to return to this point in a future work.

It was observed in \cite{Mezincescu:2010yp} that there is no choice of $a$ that avoids  anyons  in one of the levels $2$ and $3$, and that irrational spins occur for generic $a$.  Taking into account the level $4$ results,  it becomes clear that  the spectrum contains irrational spins for any choice of $a$. As we shall see, the analysis of this issue  is simpler for the superstring because supersymmetry removes the ambiguity represented by the choice of  the intercept parameter $a$.


\section{The closed ${\cal N} = 1$\, 3D Superstring}
\setcounter{equation}{0}

The action for the closed 3D  ${\cal N}=1$ GS superstring of tension $T$ is obtained from the Nambu-Goto  string action in two steps. First, we replace $d\bX$  by the supersymmetry invariant $1$-form $\Pi$ of (\ref{left-inv}) (for ${\cal N}=1$).  In other words
\be
\dot{\bX} \to \Pi_\tau =  \dot{\bX} + i \bar\Theta \Gamma \dot\Theta\, , \qquad 
{\bX}^\prime \to \Pi_\sigma =  {\bX}^\prime + i\bar\Theta\Gamma \Theta^\prime\, . 
\ee
Next,  we add to the resulting action a Wess-Zumino (WZ) term constructed from the closed, super-Poincar\'e invariant 3-form 
of (\ref{hthree1}).  Applying this prescription to the Hamiltonian form of the 3D Nambu-Goto string action (\ref{closedstring}), we find  the following `quasi-Hamiltonian' form of the  ${\cal N}=1$ 3D superstring action:
\begin{eqnarray}\label{superstring}
S[\bX,\bP;\ell,u] &=& \int \! d\tau \oint \! \frac{d\sigma}{2\pi} \left\{\Pi_\tau^\mu {\bP}_\mu     - \frac{1}{2}\ell \left[{\bP}^2 + (T\Pi_\sigma)^2\right]     
 - u \Pi_\sigma^\mu {\bP}_\mu \right. \nonumber \\
&& \left. \qquad\qquad\qquad +\ 
  iT\left( \dot{\bX}^\mu \bar\Theta \Gamma_\mu \Theta^\prime - \bX^\prime{}^\mu \bar\Theta \Gamma_\mu\dot\Theta\right) \right\} \, . 
\end{eqnarray}
By construction, this action is invariant under worldsheet diffeomorphisms, and this is equivalent 
to invariance under ``$\alpha$-symmetry''  and  ``$\beta$-symmetry'' gauge transformations that generalize (\ref{alphabetasym}). The gauge transformations of the Lagrange multiplier variables
$\ell$ and $u$ are unchanged from those of (\ref{alphabetasym}) while the canonical variables
have the gauge transformations
\begin{eqnarray}\label{alphasym1}
\delta \bX&=& \alpha\left[\bP -i\ell^{-1}\bar\Theta \Gamma
\left(\dot\Theta -u\Theta^\prime\right) \right]+ 
\beta \bX^\prime  \, , \nonumber\\
\delta \Theta &=& \alpha \ell^{-1}\left(\dot \Theta -u\Theta^\prime\right)+ \beta \Theta^\prime\, , \nonumber \\
\delta \bP &=& \left(
T^2 \alpha \Pi_\sigma + \beta\bP\right)^\prime + 2i\alpha \ell^{-1}T \left(\bar\Theta^\prime\Gamma \dot\Theta \right) \, . 
\end{eqnarray}
The term linear in $T$ in the action is the WZ term, and we have chosen its coefficient to ensure invariance of the action under the following fermionic gauge invariance (``kappa-symmetry'') with  anticommuting  Majorana spinor parameter $\kappa$:
\begin{eqnarray}
\delta_\kappa\Theta &=& \Gamma_\mu \left(\bP^\mu -T\Pi_\sigma^\mu\right) \kappa\, , \qquad
\delta_\kappa \bX^\mu = -i \bar\Theta \Gamma^\mu\delta_\kappa\Theta\, , \qquad 
\delta_\kappa\bP_\mu = 2iT \bar\Theta^\prime \Gamma_\mu \delta_\kappa\Theta\, , \nonumber\\
\delta_\kappa\ell &=&  -4i\bar\kappa \left[\dot\Theta + \left(\ell T -u\right)\Theta^\prime\right] \, , \qquad 
\delta_\kappa u = -T \delta_\kappa\ell\, . 
\end{eqnarray}
The action is $\kappa$-symmetric for either sign of $T$ but we may choose $T>0$ and then allow for either sign of the WZ term. As the two models thus obtained are equivalent we may choose the sign as given.  To verify the invariance, it is useful to use the fact that
\be
\delta_\kappa h_3 = d\delta_\kappa h_2 = -2d \left[ \Pi^\mu \left(\delta_\kappa \bar\Theta \Gamma_\mu \Theta\right)\right]\, , 
\ee
which gives $\delta_k h_2$ up to the addition of an irrelevant closed  form.  Observe that
\be
\det \left[\Gamma_\mu \left(\bP^\mu -T\Pi_\sigma^\mu\right)\right]  = - \left(\bP-T\Pi_\sigma\right)^2 \approx 0\, , 
\ee
where the symbol $\approx$ stands for  ``weak equality'' in the sense of Dirac.
This implies that  only one of the two independent components of  $\kappa$ has any effect, so that 
only one real component of   $\Theta$ can be gauged away.

The action (\ref{superstring}) is both parity invariant (for reasons explained in subsection \ref{subsec:parity}) and super-Poincar\'e invariant. The Poincar\'e Noether charges are
\begin{eqnarray}\label{noether}
{\cal P}_\mu &=& \oint\! \frac{d\sigma}{2\pi}  \left\{ {\bP}_\mu + iT \bar\Theta \Gamma_\mu \Theta^\prime\right\} \, , \nonumber \\
{\cal J}^\mu &=& \oint\! \frac{d\sigma}{2\pi} \left\{ \left[{\bX}\wedge \left({\bP}+ iT \bar\Theta\Gamma \Theta^\prime\right)\right]^\mu 
+ \frac{i}{2} \bar\Theta \Theta \left(\bP-T{\bX}^\prime\right)^\mu  \right\}\, . 
\end{eqnarray}
The supersymmetry Noether charges are 
\be
{\cal Q}^\alpha = \sqrt{2}\oint\! \frac{d\sigma}{2\pi} \left\{\left( \bP^\mu - T\Pi_\sigma^\mu \right)
\left(\Gamma_\mu \Theta\right)^\alpha - 2i T\left( \bar\Theta \Theta \right) \left(\Theta^\prime\right)^\alpha\right\}\, . 
\ee
The $\kappa$-symmetry variation of all these charges is zero on the constraint surface, i.e. weakly zero.


\subsection{Light-cone gauge} 

Light-cone gauge fixing  proceeds as for the Nambu-Goto  string but with the additional fixing of the kappa-symmetry by the relation \cite{Green:1983wt}
\be\label{fermiLCG}
\Gamma^+ \Theta =0\, , 
\ee
which implies that
\be
\Theta = \sqrt{\frac{1}{2\sqrt{2}\ p_-}}\, \left( \begin{array}{c}\theta \\ 0\end{array} \right)
\ee
for some anticommuting  worldsheet function $\theta(\tau,\sigma)$. We thus find that 
\begin{eqnarray}
\Pi_\tau^+ &=& 1\, , \qquad \Pi_\tau^- = \dot X^- + \frac{i}{2p_-}\,  \theta\dot\theta\, , \qquad \Pi_\tau^2 = \dot X\, ,  \nonumber \\
\Pi_\sigma^+ &=& 0 \, , \qquad \Pi_\sigma^- = (X^-)^\prime + \frac{i}{2p_-}\,  \theta\theta^\prime\, , \qquad \Pi_\sigma^2 = X^\prime\, . 
\end{eqnarray}

 As for the bosonic variables, it is convenient to define
\be
\bar\theta = \theta - \vartheta\, , \qquad \vartheta(\tau) = \oint \! \frac{d\sigma}{2\pi}\,  \theta\,  \, . 
\ee
There should be no confusion with the notation for a conjugate spinor as $\theta$ is not a 2-component spinor. In this notation, 
we find that the analog of (\ref{halfway}) (but without the $u=u_0 + \bar u$ split) is
\begin{eqnarray}\label{shalfway}
L &=& \dot x p  + \dot x^- p_-  + \frac{i}{2}  \vartheta \dot\vartheta  +  \oint \! \frac{d\sigma}{2\pi}\,  \left(  \dot {\bar X} \bar P   + 
\frac{i}{2}  \bar\theta \dot{\bar\theta}\right) \nonumber \\
&& + \  \frac{iT}{2 p_-} \oint \frac{d\sigma}{2\pi} \,  \bar\theta \bar\theta^\prime 
 - \oint \! \frac{d\sigma}{2\pi} \, u \left( \bar X^\prime P + \frac{i}{2} \theta \bar\theta^\prime\right) \nonumber \\
&& +\ p_-\oint \! \frac{d\sigma}{2\pi}  \left\{ \bar X^- u^\prime -  \ell\left(P_+ + \frac{1}{2p_-} \left[ P^2 + (TX^\prime)^2\right]\right) \right\}\, . 
\end{eqnarray} 
As before, $\bar X^-$ is now a Lagrange multiplier for the constraint $u'=0$, which we solve by writing $u=u_0(\tau)$. The constraint imposed by the Lagrange multiplier $\ell$ is also exactly as before, and therefore has the same solution (\ref{lapsecon}) for $P_+$.  The resulting analog of the bosonic Lagrangian (\ref{lcgLag}) is 
\begin{eqnarray}\label{sLag}
L &=& \left[ \dot x p  + {\dot x}^- p_-  + \frac{i}{2}  \vartheta \dot\vartheta + \oint \!\frac{d\sigma}{2\pi} \left\{ \dot {\bar X} \bar P  
+ \frac{i}{2}  \bar\theta \dot{\bar\theta}\right\}\right]  -H \nonumber \\
&&  - \  u_0\oint \!\frac{d\sigma}{2\pi} \left\{ \bar X^\prime \bar P + \frac{i}{2} \bar\theta \bar\theta^\prime\right\}\, , 
\end{eqnarray}
where
\begin{eqnarray}
H &=& -p_+  -  \frac{iT}{2p_-} \oint\! \frac{d\sigma}{2\pi} \, \bar\theta\bar\theta^\prime  \nonumber \\
&=&  \frac{1}{2p_-} \left[ p^2 + \oint \! \frac{d\sigma}{2\pi} \left\{ \bar P^2 + \left(T\bar X^\prime\right)^2  - iT \bar\theta \bar\theta^\prime\right\} \right]  \, . 
\end{eqnarray}
Notice that  the Hamiltonian  is no longer $-p_+$  because of the fermionic contribution from the WZ term. 

The Poincar\'e  generators in the light-cone gauge are
\begin{eqnarray}\label{Pgen}
{\cal P} &=& p \, , \qquad {\cal P}_- = p_- \, , \qquad {\cal P}_+ = -H\, , \nonumber \\
{\cal J} &=& x^- p_-  - \tau H\, , \qquad {\cal J}^+ = \tau p - x p_- \, , \nonumber \\
 {\cal J}^- &=& -x^-p -xH + \varLambda/p_-\, ,
\end{eqnarray}
exactly as for the bosonic string, except that the Hamiltonian differs and 
now
\be\label{superLambda}
\varLambda = p_-\oint \! \frac{d\sigma}{2\pi} \left[ \bar X \bar P_+  - \bar X^- \bar P \right]  
+ \frac{iT}{2}\left( \oint \! \frac{d\sigma}{2\pi}\, \bar X \bar\theta \bar\theta^\prime + \vartheta \oint \! \frac{d\sigma}{2\pi}\, \bar X \bar\theta^\prime \right)\, . 
\ee
Note the $\vartheta$-dependence in the last  term of this expression.  Note also the dependence on $\bar X^-$ in the first integral; although this is not one of the canonical variables of the gauge-fixed action, its Fourier coefficients  may again be expressed in terms of the Fourier coefficients of $(\bar X,\bar P)$. However,  in repeating this step one must  now use the relation\footnote{This corrects a minus sign error in the corresponding relation of \cite{Mezincescu:2010yp}.}
\be\label{Xminus1}
p_- \left(\bar X^-\right)^\prime + p \bar X^\prime + \frac{i}{2} \vartheta \bar\theta^\prime = - \left(\bar X^\prime \bar P + \frac{i}{2} \bar\theta\bar\theta^\prime\right) 
+ \oint \!\frac{d\sigma}{2\pi} \left(\bar X^\prime \bar P + \frac{i}{2} \bar\theta\bar\theta^\prime\right) 
\ee
which replaces (\ref{Xminus}). The relation that replaces (\ref{dotXminus}) is
\be\label{dotXminus1}
p_- D_\tau \bar X^- = \bar P_+ + i\frac{T}{2p_-} \bar\theta \bar\theta' - i \frac{T}{2p_-} \oint \frac{d\sigma}{2\pi} \, \bar\theta \bar\theta' 
\, . 
\ee

The supersymmetry charges in the light cone gauge are
\begin{eqnarray}\label{scharges}
{\cal Q}^1 &=&  \sqrt{\frac{1}{\sqrt{2}\, p_-}}\left[ p \vartheta + 
\oint \! \frac{d\sigma}{2\pi} \left(\bar P -T \bar X^\prime\right)\bar\theta \right] \, , \nonumber\\ 
{\cal Q}^2 &=&  \sqrt{\sqrt{2}\, p_-}\ \vartheta \, . 
\end{eqnarray}

Finally, the parity transformation (\ref{paritytrans1}) acts in the light-cone gauge via the transformation
\be
X \to -X\, , \qquad P\to -P\, , 
\ee
with all other canonical  variables, in particular $\theta$, being parity inert. It follows that the (classical) Hamiltonian $H$  is invariant
under parity, as expected.

\subsection{Fourier expansion} 

We Fourier expand $\bar\theta$ as 
\be\label{FourierTheta}
\bar\theta = \sum_{n=1}^\infty  \left[ e^{in\sigma} \xi_n + e^{-in\sigma} \xi_n^*\right]\, . 
\ee
With the bosonic Fourier expansions as before, the Lagrangian (\ref{sLag}) becomes
\begin{eqnarray}\label{sLag2}
L &=& \dot x p  - {\dot x}^- p_-  + \frac{i}{2}  \vartheta \dot\vartheta + i\sum_{n=1}^\infty \left[ \frac{1}{n}\left(\alpha_n^* \dot \alpha_n  + {\tilde \alpha}_n ^* {\dot {\tilde \alpha}}_n\right) + \xi_n^* \dot\xi_n \right] 
-H  \nonumber \\
&& \qquad + u_0 \sum_{n=1}^\infty \left(\alpha_n^*\alpha _n - {\tilde \alpha}^*_n {\tilde \alpha}_n + n\, \xi_n^* \xi_n \right)\, . 
\end{eqnarray}
The Hamiltonian again takes the form
\be\label{sham1}
H= \frac{1}{2p_-} \left(p^2 + {\cal M}^2\right)
\ee
but  now with 
\be
{\cal M}^2 = 2T \sum_{n=1}^\infty  \left( \alpha_n^* \alpha_n + {\tilde \alpha }_n^* {\tilde \alpha}_n + n\, \xi^*_n \xi_n \right)\, . 
\ee

The Poincar\'e charges are as in (\ref{Pgen}) but now with the different expression (\ref{superLambda}) for the Poincar\'e invariant $\Lambda$. As this involves $\bar X^-$, we must first use (\ref{Xminus1})  to express $\bar X^-$  in terms of the canonical variables, or their Fourier coefficients. The result of this computation is 
\be
-p_-\bar X^- = p \bar X + \frac{i}{2}\vartheta \bar\theta +  \sum_{n=1}^\infty \frac{i}{n} \left[ e^{in\sigma} \left(\beta_n + \gamma_n - \tilde\beta_n^*\right) - e^{-in\sigma} \left(\beta_n^* + \gamma_n^* - \tilde\beta_n\right) \right] 
\ee
where $\beta_n$ and $\tilde\beta_n$ are as they were for the bosonic string, and 
\be\label{gamman}
\gamma_n = \frac{1}{2} \sum_{m=1}^{n-1} \left(n-m\right) \xi_m\xi_{n-m} + \sum_{m>n} \left(m- \frac{n}{2}\right)\xi^*_{m-n}\xi_m\, .  
\ee
One also needs the result that
\be
\frac{i}{2} \oint \frac{d\sigma}{2\pi}\,  \bar X \bar\theta \bar\theta^\prime = \frac{1}{ \sqrt{2T}}
\sum_{n=1}^\infty \frac{i}{n} \left[\gamma_n \left(\alpha_n^* - \tilde\alpha_n\right) - 
\gamma_n^*\left(\alpha_n - \tilde\alpha_n^*\right) \right]\, . 
\ee
We then find that 
\begin{eqnarray}\label{lambdan=1}
\varLambda_+ &=&\sqrt{2T} \left[\sum_{n=1}^\infty \frac{i}{n} \alpha_n^* \left(\beta_n + \gamma_n\right)  + \frac{i}{2} \vartheta \sum_{n=1}^\infty \alpha_n^* \xi_n \right] + c.c. \, , \nonumber \\
\varLambda_- &=&   \sqrt{2T} \sum_{n=1}^\infty \frac{i}{n} \tilde\alpha_n^* \tilde\beta_n +c.c.
\end{eqnarray}

The supersymmetry charges are now
\begin{eqnarray}\label{scharges2}
{\cal Q}^1 &=&  \sqrt{\frac{1}{\sqrt{2}\, p_-}}\left[ p \vartheta + \sqrt{2T} \sum_{n=1}^\infty \left(\alpha_n \xi_n^* + \alpha_n^* \xi_n\right)\right] \, , \nonumber\\
{\cal Q}^2 &=&  \sqrt{\sqrt{2}\, p_-}\  \vartheta \, . 
\end{eqnarray}

Using (\ref{lambdan=1}) and (\ref{scharges2}), we find that the super-Poincar\'e invariant $\Omega$ of (\ref{superCasimirs}) 
takes the form 
\be
\varOmega = \varOmega_+ + \varOmega_- \,  , 
\ee
where $\varOmega_-= \varLambda_-$ and 
\be
\varOmega_+ =   \sqrt{2T} 
\sum_{n=1}^\infty \frac{i}{n}\left[ \alpha_n^*\left(\beta_n +  \gamma_n\right) - \left(\beta_n+  \gamma_n\right)^*\alpha_n  \right]\, \, . 
\ee
Note that  the anticommuting zero mode $\vartheta$,  present in $\varLambda$, cancels from $\varOmega$. 

\subsection{Quantum ${\cal N} = 1$ 3D superstring}

To quantize, we replace the bosonic variables by operators as before, and we promote the fermionic variables to operators satisfying the anti-commutation relations
\be\label{anticoms1}
\vartheta^2 = \frac{1}{2}\, , \qquad  \left\{\xi_n , \xi_n^\dagger\right\} = 1\, , 
\ee
with all other anticommutators of these variables equal to zero. 
The quantum Hamiltonian has the form (\ref{sham1}) with 
\be\label{msquared1}
{\cal M}^2 = 2T \left[ N + \tilde N + \nu\right]\, , \qquad \nu = \sum_{n=1}^\infty n \, \xi_n^\dagger \xi_n \, , 
\ee
where the bosonic level number operators $(N,\tilde N)$ are as before.  The level-matching constraint is now
\be\label{levelmatchn=1}
\tilde N = N + \nu\, ,  
\ee
which implies that
\be\label{masssqn=1}
{\cal M}^2 = 4T \left(N+\nu\right) \, , 
\ee
and hence that physical states of a given mass all appear at a particular level, given by $N+\nu$. The asymmetry in the level-matching condition  is due to the fact that the fermionic operators $\xi_n$ create right-moving modes on the string that  are super-partners to the right-moving modes, whereas the left-moving bosonic modes have no  super-partners.  In effect, the ${\cal N}=1$ GS 3D closed superstring is a 3D heterotic string.  Changing the sign of the WZ term in the action (\ref{superstring}) would lead to super-partners for the left-moving bosonic modes instead of the right-moving ones,  so there are two distinct ${\cal N}=1$ superstrings. Nevertheless, both of these potentially distinct (albeit equivalent) superstrings have exactly the same (parity preserving) 3D spectrum, so they are {\it identical} as quantum theories and we need not distinguish between them\footnote{There is nothing to prevent the strings under consideration here from self-intersecting, so we could consider a macroscopic figure-of-eight superstring in which the fermionic modes move clockwise in one loop of the ``8'' and anticlockwise in the other; this shows that  the chiral nature of the worldsheet fermions does not imply a violation of 3D parity.}.

The operator versions of  $\varOmega_\pm$ may be written as 
\be
\varOmega_+ = \sqrt{2T} \left[ \lambda + \sum_{n=1}^\infty \frac{i}{n} 
\left(\alpha_n^\dagger \gamma_n - \alpha_n\gamma_n^\dagger\right)\right] \, , \qquad 
\Omega_-\equiv \Lambda_-  =  \sqrt{2T}\, \tilde\lambda\, , 
\ee
where $\lambda$ and $\tilde\lambda$ are the operators of the bosonic string and  $\gamma_n$ is now the following operator:
\be\label{gamop}
\gamma_n =  \frac{1}{2}\sum_{m=1}^{n-1} (n-m)\xi_{m} \xi_{n-m}
+ \sum_{m>n}\left(m- \frac{n}{2}\right) \xi_{m-n}^\dagger\xi_m \, . 
\ee
The super-Poincar\'e invariant operator $\varOmega= \varOmega_+ + \varOmega_-$ is related to the Poincar\'e invariant operator 
$\varLambda$ by 
\be\label{lamom}
\varLambda = \varOmega  + \frac{1}{2\sqrt{2}}\,  i\vartheta \, \Xi \, , \qquad 
\Xi = \sqrt{4T} \sum_{n=1}^\infty \left( \alpha_n \xi_n^\dagger + \alpha_n^\dagger\xi_n\right)\, . 
\ee

The operator supercharges are
\be\label{scharges2ops}
{\cal Q}^1 =  \sqrt{\frac{1}{\sqrt{2}\, p_-}}\left[ p \vartheta + \frac{1}{\sqrt{2}} \Xi \right]   \, , \qquad 
{\cal Q}^2 =  \sqrt{\sqrt{2}\, p_-}\  \vartheta \, . 
\ee
Using the fact that 
\be\label{xisquare}
\Xi^2 = {\cal M}^2\, , 
\ee
for physical states satisfying the level-matching condition,  it is straightforward to verify that the supercharges have the expected anticommutation relations.  Although the above relation shows that the hermitian operator $\Xi$ is a square root of ${\cal M}^2$, it has zero trace in the state space to be discussed below and so is not positive; it also anticommutes with $\vartheta$. However, the hermiticity of $\Xi$ implies that ${\cal M}^2$ is positive so there exists  a positive square root  hermitian operator  ${\cal M}$; it can be defined in a basis in which ${\cal M}^2$ is diagonal by taking the positive square root of all diagonal entries.  We may then  introduce the new supercharge 
\begin{eqnarray}\label{calSdef}
{\cal S} &=& \sqrt{\sqrt{2} p_-} {\cal Q}^1 - \frac{1}{\sqrt{\sqrt{2} p_-}}
\left(p- i{\cal M}\right) {\cal Q}^2 \nonumber\\
&=& \frac{1}{\sqrt{2}} \left[i\sqrt{2} \vartheta {\cal M} + \Xi\right]\, .  
\end{eqnarray}
{}For ${\cal M}=0$ this  reduces to a factor times $\Xi$, which is (in this case)  one real linear combination of the two hermitian supercharges ${\cal Q}^\alpha$.  Otherwise, ${\cal S}$ is non-hermitian, and we may trade the two hermitian supercharges ${\cal Q}^\alpha$ for ${\cal S}$ and its hermitian conjugate.  Using the relation (\ref{xisquare}) one may verify that
\be
{\cal S}^2 =0 \, , \qquad \left\{{\cal S}, {\cal S}^\dagger\right\} = 2{\cal M}^2\, ,  
\ee
in accord with the discussion of  subsection \ref{subsec:smultiplets}. These relations are valid only when the operators act on physical states because the validity of  (\ref{xisquare}) requires the level-matching constraint.

Parity acts in the light-cone gauge of the ${\cal N}=1$ superstring in exactly the same way as it does in the bosonic theory. The parity operator $\Pi$ is again given by (\ref{parityop}); it has the property
\be
\varPi {\cal S}= - {\cal S}^\dagger\varPi \, . 
\ee
As $\varPi$ commutes with both with $N$ and $\tilde N$, as before, and trivially with $\nu$,  it commutes with the Hamiltonian. 
Since it anticommutes with $\varLambda$, this means that massive states of non-zero spin must appear in degenerate parity doublets of opposite-sign helicity. However, two such degenerate states will not appear in the same supermultiplet unless this supermultiplet has zero superspin; this is because $\varPi$ also anticommutes with $\varOmega$, so massive supermultiplets of non-zero superspin must appear in degenerate pairs of opposite sign superhelicity.

\subsubsection{Absence of  anomalies} 

The Poincar\'e charges of the ${\cal N}=1$ superstring are exactly as given in  (\ref{quantumtrans}) and (\ref{quantumLorentz})   for the bosonic string but with the Hamiltonian $H$ and Poincar\'e invariant $\varLambda$ of the superstring.  The absence of anomalies  in the Lorentz algebra is again a direct consequence of the fact that $H$ and $\varLambda$ commute, but this is now an immediate  consequence of the fact that
\be\label{fundcom}
\left[ \Xi, \varOmega\right] =0\, . 
\ee
Moreover, this relation is now the fundamental one to check because it also implies that there is no anomaly in the
commutation relation of the Lorentz charges with the supercharges. Most of the latter are just as for the superparticle; 
the only potentially problematic commutators are those which involve  ${\cal J}^-$. We should  find, for operators acting on physical states satisfying the level-matching constraint,  that
\be
\left[ {\cal J}^-, {\cal Q}^1 \right] =0\, , \qquad \left[{\cal J}^-,{\cal Q}^2\right] = -\frac{i}{\sqrt{2}} {\cal Q}^2\, . 
\ee
This can be checked directly  but it  is essentially equivalent to a check of the commutation relations 
\be
\left[ \varLambda,{\cal S}\right] = - \frac{1}{2}{\cal M} {\cal S}\, , \qquad 
\left[ \varLambda,{\cal S}^\dagger\right] =  \frac{1}{2}{\cal M}{\cal S}^\dagger\, , 
\ee
which ensure that massive supermultiplets consist of  two states differing by helicity $1/2$, and these follow directly from 
(\ref{fundcom}). 

To verify (\ref{fundcom}) we need only show that $[\Xi,\varOmega_+]=0$ since it is manifest that $\Xi$ commutes with $\varOmega_- = \varLambda_-$. As  $\Xi$ is linear  and $\Omega$ quadratic in `fermions', this  commutator contains, in principle, terms that are linear and cubic in `fermions'.  The cubic term vanishes as a consequence of the 
identity
\be
\sum_{n=1}^\infty \left( \xi_n^\dagger \gamma_n + \gamma_n^\dagger \xi_n\right) \equiv 0\, , 
\ee
which one proves by using the obvious identity
\be
\sum_{m=1}^{n-1} \xi_m \xi_{n-m} \equiv 0\, . 
\ee
To check that the term linear in `fermions' is also zero,  it  is useful to begin by establishing  the following commutation 
relations, which supplement those of  (\ref{albet}): 
\be
\left[\xi_n, \gamma_m\right] = \left(n+ \frac{m}{2}\right) \xi_{n+m}\, , \qquad
\left[ \xi_n^\dagger,\gamma_m \right] = \begin{cases}
\left(\frac{m}{2} -n\right) \xi_{m-n} & n<m \\ 0 & n=m \\ 
\left(\frac{m}{2} -n\right) \xi_{n-m}^\dagger & n>m
\end{cases}\, \quad .  
\ee


\subsubsection{Realization} 

The anticommutation relations (\ref{anticoms1}) can be partially realized by setting 
\be
\vartheta = \frac{1}{\sqrt{2}} \sigma_1 \otimes {\bI}\, , \qquad \xi_n = \sigma_2 \otimes \chi_n\, , \qquad
\xi_n^\dagger  = \sigma_2 \otimes \chi_n^\dagger\, , 
\ee
where $(\chi_n,\chi_n^\dagger)$ are a set of fermionic annihilation and creation operators:
\be
\left\{\chi_n,\chi_m^\dagger\right\} =\delta_{mn}\, .  
\ee
The operator
\be
(-1)^F = \sigma_3 \otimes \bI
\ee
anticommutes with $\vartheta$ and $\xi_n$, and hence with the supercharges\footnote{The fact that the supercharges are 3D fermions suggests an interpretation  of  $(-1)^F$ as an operator that counts spacetime fermion number modulo two. This allows us to distinguish bosonic from fermionic  {\it massless} states: recall that  this distinction survives the massless limit even though spin is not defined for massless 3D particles. The interpretation of $(-1)^F$ in its action on massive states is less clear since 
these need not be either bosons or fermions, but we pass over this point here because massive states are characterized by their relativistic helicity $s$, which we may compute directly.}. Let $|\varsigma\rangle_+$ be the pair of states ($\varsigma=\pm$) such that
\be
(-1)^F |\varsigma\rangle_+  = \varsigma|\varsigma\rangle_+ \quad (\varsigma=\pm)\, , \qquad \chi_n  |\varsigma\rangle_+  =0 
\quad (n\in \bZ^+). 
\ee
Then the doubly-degenerate  oscillator ground state (for both bosonic and fermionic operators) is
\be\label{groundn=1}
|\varsigma\rangle =  |0,\varsigma\rangle_+ \otimes |0\rangle_- \, , \qquad 
|0,\varsigma\rangle_+= |0\rangle_+ \otimes |\varsigma\rangle_+\, , 
\ee
where  $|0,\varsigma\rangle_+$ is the ground state for the right-movers  and  $|0\rangle_-$  is the ground state for the left-movers.   
The states $|\varsigma\rangle$ are annihilated by $\Xi$ and hence have zero mass.  The operator $\Xi$ is a real linear combination of the two hermitian supercharges for zero mass.  The two states of $|\varsigma\rangle$ are permuted by any other linearly-independent combination,  e.g. $\vartheta$, so they form the two states of a single massless supermultiplet. As expected, one is a boson and the other a fermion. 
Excited string states, which are  all massive,  are found by acting on the ground state $|\varsigma\rangle$  with creation operators, 
in such a way that that the  level-matching condition (\ref{levelmatchn=1}) is satisfied. We may therefore organize all physical  states according to their level $L$, with mass $M= \sqrt{4T}L$.  
In the above realization, the operators ${\cal M}$ and $\Xi$ become
\be
{\cal M} = \bI_2 \otimes {\cal M}_{\rm red}\, , \qquad \Xi = \sigma_2 \otimes \Xi_{red}
\ee
where ${\cal M}_{\rm red}$ and $\Xi_{\rm red}$ are `reduced' operators acting in the Fock space of  the operators 
$(\alpha_n,\alpha_n^\dagger)$ and $(\chi_n,\chi_n^\dagger)$.  The non-hermitian supercharge ${\cal S}$ is 
represented by
\be
{\cal S} = \frac{1}{\sqrt{2}} \left[ i\sigma_1 \otimes {\cal M}_{\rm red} + \sigma_2 \otimes \Xi_{\rm red}\right] \, . 
\ee
At a given mass level $L>0$,  for which ${\cal M}=M= \sqrt{4T}L$, we have 
\be
\left.  {\cal S} \right|_L = i\sqrt{2T}\ L \left[ \sigma_1 \otimes I_L - i \sigma_2 \otimes \eta_L \right]\, ,
\ee
where $\eta_L$ is an operator on the space of states at level $L$ that squares to the identity but has zero trace. For a given eigenvalue of $\eta_L$, we get a supermultiplet by acting with  ${\cal S}^\dagger$ on a state annihilated by ${\cal S}$ (as discussed in subsection \ref{subsec:smultiplets}) but since the eigenvalues of $\eta_L$ come in $\pm 1$ pairs, each massive level contains an even number of degenerate supermultiplets, half with $\eta_L=1$ and the other half with $\eta_L=-1$.  All massive multiplets are therefore at least quadruply degenerate.

\subsubsection{Low-lying excited states}

The first  excited states, at  level-$1$, are 
\begin{eqnarray}\label{l1n1states}
 \ |1_B,\varsigma\rangle_+\otimes|1_B\rangle_- &=& \alpha_1^\dagger |0,\varsigma\rangle_+  \otimes \tilde \alpha_1^\dagger |0\rangle_- 
\ , \nonumber\\
 |1_F,\varsigma\rangle_+\otimes|1_B\rangle_-  &=& \chi_1^\dagger |0,\varsigma\rangle_+  \otimes \tilde \alpha_1^\dagger |0\rangle_- \ ,  
\end{eqnarray}
which  gives us a total of four states at this level, and hence two ${\cal N}=1$ supermultiplets.  The level-$2$ oscillator states,  are constructed from tensor products of the `right-moving' orthonormal states 
\begin{eqnarray}\label{l2n1basis-2}
|1_B,1_B,\varsigma\rangle_+ &=& \frac{1}{\sqrt{2}}\left(\alpha_1^\dagger\right)^2  |0,\varsigma\rangle_+ \, , \qquad
|2_B,\varsigma\rangle_+ =  \frac{1}{\sqrt{2}}\alpha_2^\dagger |0,\varsigma\rangle_+ \,    \nonumber \\
|1_B,1_F,\varsigma\rangle_+ &=&  \alpha_1^\dagger\chi_1^\dagger |0,\varsigma\rangle_+ \, , \qquad  \qquad 
|2_F,\varsigma\rangle_+ =  \chi_2^\dagger |0,\varsigma\rangle_+\, ,
\end{eqnarray}
with the `left-moving'  level-$2$ states of the bosonic string
\be\label{1basis-2}
|1,1\rangle_- = \frac{1}{\sqrt{2}} \left(\tilde\alpha_1^\dagger\right)^2 |0\rangle_- \, , \qquad |2\rangle_- =  \frac{1}{\sqrt{2}}\tilde\alpha_2^\dagger |0\rangle_-\, . 
\ee
This gives us a total of 16 states, which must arrange themselves into eight  ${\cal N}=1$ supermultiplets. 

At level $3$ we need to consider the  following eight doubly-degenerate  `right-moving'  (orthonormal basis) states
\begin{eqnarray}\label{l3n1basis-3}
|1_B,1_B,1_B,\varsigma\rangle_+  &=&  \frac{1}{\sqrt{6}}\left(\alpha_1^\dagger\right)^3 |0,\varsigma\rangle_+  ,   \qquad
|1_B,2_B,\varsigma\rangle_+ =   \frac{1}{\sqrt{2}}\alpha_1^\dagger \alpha_2^\dagger |0,\varsigma\rangle_+  \, , \nonumber \\
 |3_B,\varsigma\rangle_+ &=&  \frac{1}{\sqrt{3}}\alpha_3^\dagger |0,\varsigma\rangle_+ \, , \qquad \quad \ 
 |1_F,2_F,\varsigma\rangle_+ =  \chi_1^\dagger \chi_2^\dagger|0,\varsigma\rangle_+  \, , \nonumber \\
   |1_B,2_F,\varsigma\rangle_+  &=&  \alpha_1^\dagger \chi_2^\dagger |0,\varsigma\rangle_+ \, , \qquad  \quad
   |1_B,1_B,1_F,\varsigma\rangle_+  =  \frac{1}{\sqrt{2}} \left(\alpha_1^\dagger\right)^2\chi_1^\dagger |0,\varsigma\rangle_+ \, , 
   \nonumber \\
 |2_B, 1_F,\varsigma\rangle_+ &=&  \frac{1}{\sqrt{3}}\alpha_2^\dagger\chi_1^\dagger |0,\varsigma\rangle_+ \, ,\qquad \qquad 
 |3_F,\varsigma \rangle_+ =  \chi_3^\dagger |0,\varsigma\rangle_+ \, .
\end{eqnarray}
These must be tensored with the  three  `left-moving'  level-$2$ states of the bosonic ${\cal N}=0$ string
\begin{eqnarray}\label{1basis-3}
|1,1,1\rangle_-  &=&  \frac{1}{\sqrt{6}}\left(\tilde\alpha_1^\dagger\right)^3 |0\rangle_- \, , \qquad
|1,2\rangle_- =   \frac{1}{\sqrt{2}}\tilde\alpha_1^\dagger \alpha_2^\dagger |0\rangle_- \, ,  \nonumber \\
|3\rangle_- &=&  \frac{1}{\sqrt{3}}\tilde\alpha_3^\dagger |0\rangle_- \, . 
\end{eqnarray}
This gives us a total of 48 states at level $3$, and hence 24 ${\cal N}=1$ supermultiplets.

\subsubsection{Superhelicities} 

In order to determine the spectrum of superhelicities,  we must find the eigenvalues  of the operator $\varOmega$, which we can do 
by finding those of $\varOmega_+$ since the eigenvalues of $\varOmega_-=\varLambda_-$ have already been computed (to the level considered here). There will be a double degeneracy in the eigenvalues of  $\varOmega$ due to the independence of this operator on the  fermion zero mode $\vartheta$; this is the degeneracy implied by ${\cal N}=1$ supersymmetry. As discussed in the previous subsection, there is a further double-degeneracy in massive levels, so the number of potentially distinct eigenvalues of $\varOmega$ at a given mass level is only a quarter of the number of states at that level.  A further simplifying feature is that $\varOmega_+$ is  a Grassmann even operator that does not mix states of different Grassmann parity (as determined by the Grassman parities of the operators used to construct the states, ignoring  the Grassmann parity of the ground states).

The operator $\varOmega$ annihilates the two massless ground states, which form a massless supermultiplet of ${\cal N}=1$ 3D supersymmetry comprising one boson and one fermion (recall that spin is not defined  for massless particles). Potentially, these could be identified as a dilaton and dilatino.  The operator $\varOmega$ also annihilates all  level-$1$ states, which implies that the four states at this level yield two copies of the ${\cal N}=1$ semion supermultiplet with helicities  ($\frac{1}{4},  -\frac{1}{4}$); this is the supermultiplet one gets by quantizing the ${\cal N}=1$ massive superparticle \cite{Mezincescu:2010yq}.  At level $2$, there are 16 states and so eight eigenvalues of $\varOmega$. To compute them,  we need to consider the $4\times 4$ matrix that results from the action of  $\varOmega_+$  in the space spanned by the  basis states (\ref{l2n1basis-2}). We may write this operator  in the form
\be
\varOmega_+ = \sqrt{2T}\sum_{n=2}^\infty \omega_n \, , 
\ee
where the operator $\omega_n$, which generalizes the operator $\lambda_n$ introduced earlier in (\ref{lambdasum}), 
annihilates all states with $N+\nu<n$ but not all those with $N+\nu \ge n$. At level $2$ we need only 
\be
\omega_2 =  \frac{3i}{4} \left[ \left(\alpha_1^\dagger\right)^2 \alpha_2 - \alpha_2^\dagger \alpha_1^2\right] 
+ \frac{3i}{2}\left[ \alpha_1^\dagger\xi_1^\dagger\xi_2 - \xi_2^\dagger\xi_1\alpha_1\right] \, . 
\ee
Each of the two terms in this expression contributes to only one $2\times 2$ block of the $4\times 4$ matrix, which is therefore block-diagonal. The first term is $\lambda_2$ and we showed earlier that this gives the $2\times 2$ matrix $(3/2)\sigma_2$. The second block, coming from the second term, also turns out to be $(3/2)\sigma_2$, so the eigenvalues of $\varOmega_+$ at level $2$ are 
$\sqrt{2T}(\frac{3}{2}, \frac{3}{2},-\frac{3}{2},-\frac{3}{2})$. The eigenvalues of  $\varOmega_- = \varLambda_-$ are  $\sqrt{2T}(\frac{3}{2},-\frac{3}{2})$ so the eight eigenvalues of $\varOmega$ at level $2$ are $\sqrt{2T}(0,0,0,0, 3,3,-3,-3)$.  To get the superhelicities $s_1$ (recall that the subscript here indicates the number ${\cal N}$ of supersymmetries)   we have to divide by the level-$2$ mass, which is $2\sqrt{2T}$; this gives
\be
s_1= \left(-\frac{3}{2},-\frac{3}{2}, 0,0,0,0, \frac{3}{2},\frac{3}{2}\right)\, . 
\ee
The supermultiplets  with  superhelicity $s_1=\pm \frac{3}{2}$ have  helicities $s=(\pm\frac{7}{4}, \pm \frac{5}{4}) $.  

These level-$1$ and level-$2$ results, which show that semions are present in the spectrum of the 3D ${\cal N}=1$  superstring, were announced in \cite{Mezincescu:2010yp}. Here we continue the analysis to the next level.  At level 3 we need 
\begin{eqnarray}
\omega_3 &=& \frac{7i}{6} \left(\alpha_1^\dagger \alpha_2^\dagger \alpha_3 - \alpha_3^\dagger \alpha_1 \alpha_2\right)   \\
&&+ \, i \left[ \frac{5}{2}\left(\alpha_1^\dagger\xi_2^\dagger\xi_3 - \xi_3^\dagger\xi_2\alpha_1\right) +\left(\alpha_2^\dagger
 \xi_1^\dagger\xi_3 - \xi_3^\dagger\xi_1\alpha_2\right) - \frac{1}{6}\left(\alpha_3^\dagger\xi_1\xi_2 - \xi_2^\dagger\xi_1^\dagger\alpha_3\right)\right] . \nonumber
\end{eqnarray}
In the level-$3$ basis (\ref{l3n1basis-3}), $\omega_2+ \omega_3$  is a block-diagonal $8\times 8$ matrix with the following two $4\times 4$ blocks:
\be
i\left( \begin{array}{cccc} 0 & -\frac{3\sqrt{3}}{2} &0  & 0\\ \frac{3\sqrt{3}}{2} & 0 & - \frac{7}{\sqrt{6}} & 0 \\ 0& \frac{7}{\sqrt{6}} &0& -\frac{1}{2\sqrt{3}} \\
0 & 0&\frac{1}{2\sqrt{3}}& 0  \end{array} \right)\, , \qquad 
i\left( \begin{array}{cccc} 0 & +\frac{3}{\sqrt{2}} &0  & -\frac{5}{2}\\ -\frac{3}{\sqrt{2}} & 0 & - \frac{3}{2} & 0 \\ 0& \frac{3}{2} &0 &-\sqrt{2} \\
\frac{5}{2} & 0&\sqrt{2} & 0  \end{array} \right).
\ee
The eigenvalues of these matrices are,  respectively,
\be\label{Omega3}
\pm\sqrt{\frac{15}{2} \pm \frac{ 9\sqrt{11}}{4}} \qquad \mbox{and}   \qquad  \pm \sqrt{ 3 \pm \frac{ 3\sqrt{7}} {4} } \, ,
\ee
which shows that at the level 3 there are supermultiplets of  irrational superhelicity.

\section{The closed ${\cal N} = 2$\, 3D Superstring}
\setcounter{equation}{0}

As for ${\cal N}=1$, the ${\cal N}=2$ GS superstring action can be obtained from the bosonic string action in two steps. First, we make the  replacement
\be
\dot\bX \to \Pi_\tau =  \dot{\bX} + i \bar\Theta_a \Gamma \dot\Theta_a\, , \qquad 
\bX^\prime \to \Pi_\sigma =  {\bX}^\prime + i\bar\Theta_a\Gamma \Theta_a^\prime\, .
\ee
Next,  we add to the resulting action a Wess-Zumino (WZ) term constructed from the closed, super-Poincar\'e invariant 3-form
$h_3$ given in (\ref{hthree}).  This can be written as $h_3=dh_2$ for $h_2$  as given in (\ref{htwo}), and the integral of $h_2$ gives us the required WZ term. These considerations lead to the following quasi-Hamiltonian form of the ${\cal N}=2$ superstring action:
\begin{eqnarray}\label{2superstring}
S[\bX,\bP,\Theta_a;\ell,u] &= & \!\! \int \! d\tau \oint \! \frac{d\sigma}{2\pi} \left\{\Pi_\tau^\mu {\bP}_\mu     - \frac{1}{2}\ell \left[{\bP}^2 + (T\Pi_\sigma)^2\right]    - u\,  \Pi_\sigma^\mu {\bP}_\mu \right. \nonumber \\
 &&+\  iT\left[\left( \dot{\bX}^\mu +\frac{i}{2}{\bar \Theta}_a\Gamma^\mu{\dot \Theta}_a\right) \left( \bar\Theta_1 \Gamma_\mu \Theta'_1 - \bar\Theta_2 \Gamma_\mu \Theta'_2\right)\right.   \\
 &&  \left. \left. \qquad -  \left( {\bX}'^\mu +\frac{i}{2}{\bar \Theta}_a\Gamma^\mu{ \Theta'}_a\right) \left( \bar\Theta_1 \Gamma_\mu {\dot \Theta}_1 - \bar\Theta_2 \Gamma_\mu {\dot\Theta}_2\right)\right] \right\} \, . \nonumber
\end{eqnarray}
This action has $\alpha$-symmetry and $\beta$-symmetry gauge invariances that generalize those of the ${\cal N}=1$ 3D superstring.  The  transformations of the Lagrange multiplier variables are unchanged while those of the canonical variables are 
\begin{eqnarray}\label{alphasym2}
\delta \bX&=& \alpha\left[\bP -i\ell^{-1}\bar\Theta_a \Gamma
\left(\dot\Theta_a -u\Theta_a^\prime\right) \right]+ 
\beta \bX^\prime  \, , \nonumber\\
\delta \Theta_a &=& \alpha \ell^{-1}\left(\dot \Theta_a -u\Theta_a^\prime\right)+ 
\beta \Theta_a^\prime\, , \nonumber \\
\delta \bP &=& \left(
T^2 \alpha \Pi_\sigma + \beta\bP\right)^\prime + 2i\alpha \ell^{-1}T \left(\bar\Theta_1^\prime\Gamma \dot\Theta_1 - \bar\Theta_2^\prime\Gamma\dot\Theta_2 \right) \, . 
\end{eqnarray}

The term linear in $T$ in the action (\ref{2superstring}) is the WZ term, and we have chosen its coefficient to ensure invariance under the following ``$\kappa$-symmetry'' gauge transformation  with  anticommuting  Majorana spinor parameters $\kappa_a$:
\begin{eqnarray}
\delta_\kappa \bX^\mu &=& -i \bar\Theta_a \Gamma^\mu\delta_\kappa\Theta_a\, , \qquad 
\delta_\kappa\bP_\mu = 2iT\left( \bar\Theta_1^\prime \Gamma_\mu \delta_\kappa\Theta_1-  \bar\Theta_2^\prime \Gamma_\mu \delta_\kappa\Theta_2\right)\, , \nonumber \\
\delta_\kappa\Theta_1 &=& \Gamma_\mu \left(\bP^\mu -T\Pi_\sigma^\mu\right) \kappa_1\,, \qquad
\delta_\kappa\Theta_2= \Gamma_\mu \left(\bP^\mu +T\Pi_\sigma^\mu\right) \kappa_2\,, \nonumber \\
\delta_\kappa\ell &=&  -4i\bar\kappa_1 \left[\dot\Theta_1 + \left(\ell T -u\right)\Theta_1^\prime\right]  -4i\bar\kappa_2 \left[\dot\Theta_2 + \left(-\ell T -u\right)\Theta_2^\prime\right] \, , \nonumber \\
\qquad \delta_\kappa u &=& -T \left(\delta_{\kappa_1}\ell - \delta_{\kappa_2}\ell\right)\, . 
\end{eqnarray}
Because of the relative minus sign in the WZ term, its overall sign can be changed by the field redefinition $\Theta_1 \leftrightarrow \Theta_2$, so we may choose $T>0$.  To verify the $\kappa$-symmetry, it is useful to use the fact that
\be
\delta_\kappa h_3 = d\delta_\kappa h_2 = -2d \left[ \Pi^\mu \left(\delta_\kappa \bar\Theta_1 \Gamma_\mu \Theta_1 - 
\delta_\kappa \bar\Theta_2 \Gamma_\mu \Theta_2\right)\right]\, , 
\ee
which gives $\delta_k h_2$ up to the addition of an irrelevant closed  form.  
Observe that 
\be
\det \left[\Gamma_\mu \left(\bP^\mu \mp T\Pi_\sigma^\mu\right)\right]  = - \left(\bP\mp T\Pi_\sigma\right)^2 \approx 0\, . 
\ee
As for the $ N= 1$ superstring, this implies that  only one of the two independent components of each  $\kappa_a$ has any effect, so that 
only one real component of  each $\Theta_a$ can be gauged away.

As for rigid symmetries, the action (\ref{2superstring}) is invariant under the parity transformation of (\ref{paritytrans2}).  It is also super-Poincar\'e invariant, by construction. 
The Poincar\'e Noether charges  are
\begin{eqnarray}\label{noether}
{\cal P}_\mu &=& \oint\! \frac{d\sigma}{2\pi}  \left\{ {\bP}_\mu + iT \left[\bar\Theta_1 \Gamma_\mu \Theta_1^\prime - \bar\Theta_2 \Gamma_\mu \Theta_2^\prime \right]\right\} \, , \nonumber \\
{\cal J}^\mu &=& \oint\! \frac{d\sigma}{2\pi} \left\{ \left[{\bX}\wedge \left({\bP}+  iT \left(\bar\Theta_1\Gamma \Theta_1^\prime - \bar\Theta_2\Gamma \Theta_2^\prime\right)\right)\right]^\mu  \right. \nonumber \\
&&  \quad \qquad \  
+ \frac{i}{2} \bar\Theta_1 \Theta_1 \left(\bP-T{\bX}^\prime\right)^\mu   +  \frac{i}{2} \bar\Theta_2 \Theta_2 \left(\bP+T{\bX}^\prime\right)^\mu \, \nonumber \\
&& \left.  \quad \qquad -\ (T/2) \left(\bar\Theta_2 \Gamma^\mu\Theta^\prime_2 \bar\Theta_1 \Theta_1 -  \bar\Theta_1 \Gamma^\mu\Theta^\prime_1 \bar\Theta_2 \Theta_2\right)\right\}.
\end{eqnarray}
The supersymmetry Noether charges are 
\begin{eqnarray}
{\cal Q}_1^\alpha &=& \sqrt{2}\oint\! \frac{d\sigma}{2\pi} \left\{ \left(\bP^\mu -T\Pi_\sigma \right) \left(\Gamma^\mu \Theta_1\right)^\alpha
-2iT \left(\bar\Theta_1\Theta_1\right) \Theta_1^\prime \right\}\, , \nonumber \\
{\cal Q}_2^\alpha &=& \sqrt{2}\oint\! \frac{d\sigma}{2\pi} \left\{ \left(\bP^\mu +T\Pi_\sigma \right) \left(\Gamma^\mu \Theta_2\right)^\alpha
+ 2iT \left(\bar\Theta_2\Theta_2\right) \Theta_2^\prime \right\}\, . 
\end{eqnarray}

\subsection{Light-cone gauge}

The light-cone gauge fixing proceeds as for the ${\cal N} = 1$ superstring but with the additional fixing of the larger  kappa-symmetry gauge invariance by the condition 
\be
\Gamma^+ \Theta_a =0\, , \qquad a=1,2. 
\ee
In this gauge, 
\be
\Theta_a = \sqrt{\frac{1}{2\sqrt{2}\ p_-}}\, \left( \begin{array}{c}\theta_a \cr 0\end{array} \right)
\ee
for some anticommuting  worldsheet functions $\theta_a(\tau,\sigma)$. We thus find that 
\begin{eqnarray}
\Pi_\tau^+ &=& 1\, , \qquad \Pi_\tau^- = \dot X^- + \frac{i}{2p_-}\,  \theta_a\dot\theta_a\, , \qquad \Pi_\tau^2 = \dot X \, , \nonumber \\
\Pi_\sigma^+ &=& 0 \, , \qquad \Pi_\sigma^- = (X^-)^\prime + \frac{i}{2p_-}\,  \theta_a\theta_a^\prime\, , \qquad \Pi_\sigma^2 = X^\prime\, . 
\end{eqnarray}

Again it is convenient, it is convenient to define
\be
\bar\theta_a = \theta_a - \vartheta_a\, , \qquad \vartheta_a(\tau) = \oint \! \frac{d\sigma}{2\pi}\,  \theta_a\,  \, . 
\ee
Again, there should be no confusion with the notation for a conjugate spinor as the $\theta_a$ are not 2-component spinors. In this notation, 
we find that the analog of (\ref{halfway}) (but without the $u=\bar u + u_0$ split) is 
\begin{eqnarray}\label{shalfway}
L &=& \dot x p  + \dot x^- p_-  + \frac{i}{2}  \vartheta_a \dot\vartheta_a  +  \oint \! \frac{d\sigma}{2\pi}\,  \left\{  \dot {\bar X} \bar P   + 
\frac{i}{2}  \bar\theta_a {\dot{\bar\theta}}_a\right\} \nonumber \\
&& + \  \frac{iT}{2 p_-} \oint \!\frac{d\sigma}{2\pi} \,  \left(\bar\theta_1 \bar\theta_1^\prime - \bar\theta_2 \bar\theta_2^\prime \right)
 - \oint \! \frac{d\sigma}{2\pi} \, u \left[ \bar X^\prime P + \frac{i}{2} \theta_a \bar\theta_a^\prime\right] \nonumber \\
&& +\  p_-\oint \! \frac{d\sigma}{2\pi}  \left\{ \bar X^- u^\prime -  \ell\left(P_+ + \frac{1}{2p_-} \left[ P^2 + (TX^\prime)^2\right]\right) \right\}\, . 
\end{eqnarray} 
As before, $\bar X^-$ is now a Lagrange multiplier for the constraint $u'=0$, which we solve by writing $u=u_0(\tau)$. The constraint imposed by the Lagrange multiplier $\ell$ is also exactly as before, and therefore has the same solution (\ref{lapsecon}) for $P_+$.  The resulting analog of the bosonic Lagrangian (\ref{lcgLag}) is 
\begin{eqnarray}\label{sLag1}
L &=& \left[ \dot x p  + {\dot x}^- p_-  + \frac{i}{2}  \vartheta_a \dot\vartheta_a + \oint \!\frac{d\sigma}{2\pi} \left\{ \dot {\bar X} \bar P  
+ \frac{i}{2}  \bar\theta_a {\dot {\bar\theta}}_a\right\}\right]  -H \nonumber \\
&&  - \  u_0 \oint \!\frac{d\sigma}{2\pi} \left\{ \bar X^\prime \bar P + \frac{i}{2} \bar\theta_a \bar\theta_a^\prime\right\}\, , 
\end{eqnarray}
where
\begin{eqnarray}\label{HamLCG}
H &=& -p_+  -  \frac{iT}{2p_-} \oint\! \frac{d\sigma}{2\pi} \,\left(\bar\theta_1\bar\theta_1^\prime - \bar\theta_2\bar\theta_2^\prime\right)  \nonumber \\
&=&  \frac{1}{2p_-} \left[ p^2 + \oint \! \frac{d\sigma}{2\pi} \left\{ \bar P^2 + \left(T\bar X^\prime\right)^2  - iT \left(\bar\theta_1\bar\theta_1^\prime - \bar\theta_2\bar\theta_2^\prime\right)\right\} \right]  \, . 
\end{eqnarray}
As for ${\cal N}=1$, the Hamiltonian  is not equal to  $-p_+$  because it gets a fermionic contribution from the WZ term. 

The Poincar\'e  generators in the light-cone gauge are
\begin{eqnarray}\label{n=2pgenerators}
{\cal P} &=& p \, , \qquad {\cal P}_- = p_- \, , \qquad {\cal P}_+ = -H\, , \nonumber \\
{\cal J} &=& x^- p_-  +  \tau H\, , \qquad {\cal J}^+ = \tau p - x p_- \, , \nonumber \\
 {\cal J}^- &=& -x^-p -xH + \varLambda/p_-\, ,
\end{eqnarray}
exactly as for the ${\cal N}=1$ superstring except that the Hamiltonian differs and now
now
\begin{eqnarray}
\varLambda &=& p_- \oint \! \frac{d\sigma}{2\pi} \left( \bar X \bar P_+  - \bar X^- \bar P \right)  +
\frac{iT}{2} \oint \! \frac{d\sigma}{2\pi}\, \bar X \left(\bar\theta_1 \bar\theta_1^\prime - \bar\theta_2 \bar\theta_2^\prime\right) \nonumber \\
&& \qquad \qquad 
+\ \frac{iT}{2}\left( \vartheta_1 \oint \! \frac{d\sigma}{2\pi}\, \bar X \bar\theta_1^\prime - \vartheta_2 \oint \! \frac{d\sigma}{2\pi}\, 
\bar X \bar\theta_2^\prime\right) \, . 
\end{eqnarray}
Note the $\vartheta_a$-dependence of this expression. The Fourier coefficients of  $\bar X^-$  may again be expressed in terms of the Fourier coefficients of $(\bar X,\bar P)$, but  in repeating this step we should now use the ${\cal N}=2$ relation 
\be\label{sXminus2}
p_- \left(\bar X^-\right)^\prime + p \bar X^\prime + \frac{i}{2}\vartheta_a \bar\theta_a^\prime = - \left(\bar X^\prime \bar P + \frac{i}{2} \bar\theta_a\bar\theta_a^\prime\right) 
+ \oint \!\frac{d\sigma}{2\pi} \left(\bar X^\prime \bar P + \frac{i}{2} \bar\theta_a\bar\theta_a^\prime\right) \, , 
\ee
which replaces (\ref{Xminus1}).  The relation that replaces (\ref{dotXminus1}) is
\be\label{dotXminus2}
p_- D_\tau \bar X^- = \bar P_+ + i\frac{T}{2p_-}\left(\bar\theta_1\bar\theta_1^\prime - \bar\theta_2\bar\theta_2^\prime\right)  
- i \frac{T}{2p_-} \oint \frac{d\sigma}{2\pi} \left(\bar\theta_1 \bar\theta_1^\prime - \bar\theta_2 \bar\theta_2^\prime\right)
\, . 
\ee

The supersymmetry charges in the light-cone gauge are
\begin{eqnarray}\label{scharges}
{\cal Q}_1^1 &=&  \sqrt{\frac{1}{\sqrt{2}\, p_-}}\left[ p \vartheta_1 + \oint \! \frac{d\sigma}{2\pi} \left(\bar P -T \bar X^\prime\right)\bar\theta_1 \right]\, ,  \nonumber\\
{\cal Q}_1^2 &=&  \sqrt{\sqrt{2}\, p_-}\ \vartheta_1 \, , 
\end{eqnarray}
and 
\begin{eqnarray}\label{scharges}
{\cal Q}_2^1 &=&  \sqrt{\frac{1}{\sqrt{2}\, p_-}}\left[ p \vartheta_2 +  \oint \! \frac{d\sigma}{2\pi} \left(\bar P +T \bar X^\prime\right)\bar\theta_2 \right] \, , \nonumber\\
{\cal Q}_2^2 &=&  \sqrt{\sqrt{2}\, p_-}\ \vartheta_2 \, . 
\end{eqnarray}

Finally, parity acts in the light-cone gauge via the transformation
\be
X\to -X \, , \qquad P\to -P \, , \qquad \theta_2 \to -\theta_2\, , 
\ee
with all other canonical variables  being parity inert.  The light-cone gauge Hamiltonian (\ref{HamLCG}) is parity  invariant, as expected.

\subsection{Fourier expansion}

We Fourier expand the $\bar\theta_a$ as
\be
\bar\theta_1 = \sum_{n=1}^\infty  \left[ e^{in\sigma} \xi_n + e^{-in\sigma} \xi_n^*\right]\, , \qquad 
\bar\theta_2 = \sum_{n=1}^\infty  \left[ e^{in\sigma} {\tilde\xi}^*_n + e^{-in\sigma} {\tilde\xi}_n\right]\, . 
\ee
With the bosonic Fourier expansions as before, the Lagrangian (\ref{sLag1}) becomes
\begin{eqnarray}\label{sLag2}
L &=& \dot x p  - {\dot x}^- p_-  + \frac{i}{2}  \vartheta_a \dot\vartheta_a + i\sum_{n=1}^\infty \left[ \frac{1}{n}\left(\alpha_n^* \dot \alpha_n  + {\tilde \alpha}_n ^* {\dot {\tilde \alpha}}_n\right) + \xi_n^* \dot\xi_n + {\tilde\xi}_n^* \dot{\tilde\xi}_n\right] 
-H  \nonumber \\
&& \qquad +\  u_0 \sum_{n=1}^\infty \left[\alpha_n^*\alpha _n - {\tilde \alpha}^*_n {\tilde \alpha}_n + n\left(\xi_n^* \xi_n - {\tilde\xi}_n^* {\tilde\xi}_n\right)\right]\, . 
\end{eqnarray}
The Hamiltonian again takes the form
\be\label{sham}
H= \frac{1}{2p_-} \left(p^2 + {\cal M}^2\right)
\ee
but  now with 
\be
{\cal M}^2 = 2T \sum_{n=1}^\infty  \left[ \alpha^* \alpha + {\tilde \alpha }_n^* {\tilde \alpha}_n + n\left(\xi^*_n \xi_n + {\tilde\xi}_n^* {\tilde\xi}_n \right)\right]\, . 
\ee

Similarly, the Poincar\'e charges are as in (\ref{n=2pgenerators}) with $\varLambda= \varLambda_++ \varLambda_+$, with
\begin{eqnarray}
\varLambda_+ &=& \varOmega_+ + \sqrt{\frac{T}{2}} i\vartheta_1 \sum_{n=1}^\infty 
\left(\alpha_n^* \xi_n + \alpha_n \xi_n^*\right)\, , \nonumber \\
\varLambda_- &=& \varOmega_- + \sqrt{\frac{T}{2}} i\vartheta_2 \sum_{n=1}^\infty 
\left(\tilde\alpha_n^* \tilde\xi_n + \tilde\alpha_n \tilde\xi_n^*\right)\, , 
\end{eqnarray}
where $\varOmega_\pm$, which sum to the ${\cal N}=2$ super-Poincar\'e invariant $\varOmega$, are given by 
\begin{eqnarray}
\varOmega_+  &=&   \sqrt{2T} \left[\lambda + 
\sum_{n=1}^\infty \frac{i}{n} \left(\alpha_n^* \gamma_n - \alpha_n \gamma_n^* \right) \right]\, , \nonumber \\
\varOmega_-  &=&   \sqrt{2T} \left[\tilde\lambda + 
\sum_{n=1}^\infty \frac{i}{n} \left(\tilde\alpha_n^* \tilde\gamma_n - \tilde\alpha_n \tilde\gamma_n^* \right) \right]\, . 
\end{eqnarray}
In these expressions, the quantities $\lambda$ and $\tilde\lambda$ are as given in (\ref{littlelambdas}) for the bosonic string and  $\gamma_n$ is as given in 
 (\ref{gamman}) for the ${\cal N}=1$ string, with a formally identical expression for $\tilde\gamma_n$ in terms of the `left-moving' canonical variables. Note that  the fermionic zero modes  $\vartheta_a$ cancel from $\varOmega_\pm$.

The supersymmetry charges are
\begin{eqnarray}\label{schargesn=1}
{\cal Q}_1^1 &=&  \sqrt{\frac{1}{\sqrt{2}\, p_-}}\left[ p \vartheta_1 + \sqrt{2T} \sum_{n=1}^\infty \left\{\alpha_n \xi_n^* + \alpha_n^* \xi_n\right\} \right] \, , \nonumber\\
{\cal Q}_1^2 &=&  \sqrt{\sqrt{2}\, p_-}\  \vartheta_1 \, , 
\end{eqnarray}
and
\begin{eqnarray}\label{schargesn=2}
{\cal Q}_2^1 &=&  \sqrt{\frac{1}{\sqrt{2}\, p_-}}\left[ p \vartheta_2 + \sqrt{2T} \sum_{n=1}^\infty \left\{\tilde\alpha_n \tilde\xi_n^* + \tilde\alpha_n^* \tilde\xi_n\right\} \right] \, , \nonumber\\
{\cal Q}_2^2 &=&  \sqrt{\sqrt{2}\, p_-}\  \vartheta_2 \, . 
\end{eqnarray}
The upper number  is the value of the spinor index $\alpha$, and the lower number is  the value of the supersymmetry-number index $a$.

Parity now acts via the transformations
\begin{eqnarray}
x &\to& -x\, , \qquad p\to -p\, , \qquad \vartheta_2 \to -\vartheta_2\, , \nonumber\\
\alpha_n &\to& - \alpha_n \, , \qquad \tilde\alpha_n \to -\tilde\alpha_n \, , \qquad \tilde\xi_n \to - \tilde \xi_n\, . 
\end{eqnarray}
The asymmetry in the action on the fermi modes originates in the relative minus sign in the 
$\Theta_a$ transformation of (\ref{paritytrans2}).

\subsection {Quantum ${\cal N} = 2$ 3D superstring}

To quantize, we replace the bosonic variables by operators as before, and we promote the fermionic variables to operators satisfying the anti-commutation relations
\be\label{anticoms}
\left\{\vartheta_a, \vartheta_b\right\} =  \delta_{ab}\, , \qquad  \left\{\xi_n , \xi_n^\dagger\right\} = 1, \qquad \left\{\tilde\xi_n , \tilde\xi_n^\dagger\right\} = 1\, , 
\ee
with all other anticommutators of these variables equal to zero. 
The quantum Hamiltonian has the form (\ref{sham}) with 
\be\label{msquaredn=2}
{\cal M}^2 = 2T \left[ N + \tilde N + \nu + \tilde\nu\right]\, , \qquad \nu = \sum_{n=1}^\infty n \, \xi_n^\dagger \xi_n, \qquad \tilde\nu = \sum_{n=1}^\infty n \, \tilde\xi_n^\dagger \tilde\xi_n \, ,
\ee
where the bosonic level number operators $(N,\tilde N)$ are as before.  The level-matching constraint, on the eigenvalues of these operators, is now
\be\label{levelmatchn=2}
\tilde N + \tilde\nu = N + \nu\, , 
\ee
and we may use this to rewrite the mass-squared at level $L=N+\nu$ as 
\be\label{lmasssqmatch}
\left. {\cal M}^2\right|_L = 4T L \, , \qquad L=  N+\nu \, . 
\ee

The quantum supersymmetry charges are obtained from the classical charges (\ref{schargesn=1}) and (\ref{schargesn=1}) in the  usual way.  The result is
\begin{eqnarray}\label{schargesn=1}
{\cal Q}_1^1 &=&  \sqrt{\frac{1}{\sqrt{2}\, p_-}}\left[ p \vartheta_1 + \frac{1}{\sqrt{2}} \Xi\right] \, , \qquad
{\cal Q}_1^2 =  \sqrt{\sqrt{2}\, p_-}\  \vartheta_1 \, , \nonumber\\
{\cal Q}_2^1 &=&  \sqrt{\frac{1}{\sqrt{2}\, p_-}}\left[ p \vartheta_2 + \frac{1}{\sqrt{2}} \tilde\Xi\right] 
\, , \qquad  {\cal Q}_2^2 =  \sqrt{\sqrt{2}\, p_-}\  \vartheta_2 \, , 
\end{eqnarray}
where 
\be
\Xi= \sqrt{4T}\, \sum_{n=1}^\infty \left( \alpha_n \xi_n^\dagger + \alpha_n^\dagger \xi_n \right)\, , 
\qquad 
\tilde\Xi= \sqrt{4T}\, \sum_{n=1}^\infty \left(\tilde \alpha_n \tilde\xi_n^\dagger + \tilde\alpha_n^\dagger \tilde\xi_n \right)\, .
\ee
The operators  $\Xi$ and $\tilde\Xi$ also appear in the relation between the quantum operators $\varOmega_\pm$ and $\varLambda_\pm$:
\be
\varLambda_+ = \varOmega_+ + \frac{i}{2\sqrt{2}} \vartheta_1 \Xi\, , \qquad 
\varLambda_- = \varOmega_- + \frac{i}{2\sqrt{2}} \vartheta_2 \tilde\Xi\, . 
\ee
When these operators act on physical states satisfying the level-matching condition (\ref{levelmatchn=2}), they satisfy
\be\label{xicomrel}
\Xi^2= {\cal M}^2 = \tilde\Xi^2 \,, \qquad \left\{\Xi,\tilde\Xi\right\} =0\, . 
\ee
For the reasons explained earlier for the ${\cal N}=1$ superstring, the absence of super-Poincar\'e anomalies is a consquence of the fact 
\be
\left[\Xi,\varOmega_+\right] =0 \, , \qquad \left[\tilde\Xi,\varOmega_-\right] =0 \, . 
\ee
The  calculations needed to verify these commutation relations are also the same as those sketched earlier for the
 ${\cal N}=1$ superstring. 

 As for the ${\cal N}=1$ superstring, it is convenient to consider the supercharges
\be
{\cal S}_a = \sqrt{\sqrt{2}\, p_-}\ {\cal Q}_a^1 - \frac{1}{\sqrt{\sqrt{2}\, p_-}}\left(p-i{\cal M}\right) {\cal Q}^2_a\, , 
\ee
where ${\cal M}$ is the positive square root of ${\cal M}$.  This gives
\be
{\cal S}_1 =   i{\cal M} \vartheta_1 + \frac{1}{\sqrt{2}} \Xi \, , \qquad 
{\cal S}_2 =  i{\cal M} \vartheta_2 + \frac{1}{\sqrt{2}} \tilde\Xi \, .
\ee
Using (\ref{xicomrel}), it is straightforward to verify that 
\be
\{{\cal S}_a ,{\cal S}_b \} =  0 \, , \qquad \left\{ {\cal S}_a, {\cal S}_b^\dagger\right\} = 2\delta_{ab}\,  {\cal M}^2\, . 
\ee
The operators ${\cal S}_a$ again commute with the operator version of $\varOmega$, but 
\be
\left[\varLambda, {\cal S}_a\right] = - \frac{1}{2} {\cal M} {\cal S}_a\, , \qquad 
\left[\varLambda, {\cal S}_a^\dagger \right] = \frac{1}{2} {\cal M} {\cal S}_a^\dagger\, .
\ee

The parity operator in the light-cone gauge takes the form 
\be
\varPi_{{\cal N}=2} = \varPi(-1)^{F_L}
\ee
where $\varPi$ is the parity operator (\ref{parityop}) of the bosonic string and  ${\cal N}=1$ superstring, and the operator $(-1)^{F_L}$ anticommutes with $\vartheta_2$ and all $\tilde\xi_n$ but commutes with all other canonical variables.  As this operator anticommutes with both $\varLambda_\pm$ and $\varOmega_\pm$, both helicity and superhelicity eigenstates must appear in parity doublets of opposite sign eigenvalues.

We may similarly define an operator $(-1)^{F_R}$ that anticommutes with $\vartheta_1$ and all $\xi_n$ but commutes with all other canonical variables. The operator
\be
(-1)^F = (-1)^{F_L} (-1)^{F_R}
\ee
anticommutes with all fermionic canonical variables but commutes with all the bosonic canonical variables. As a consequence it anticommutes with all components of the supercharges ${\cal Q}_a$, so the action of one of these charges on an eigenstate of 
$(-1)^F$ yields another eigenstate of $(-1)^F$ but with opposite sign eigenvalue.

\subsubsection{Realization}

The canonical anticommutation relations (\ref{anticoms}) can be partially realized by setting
\begin{eqnarray}\label{partialrep}
\sqrt{2} \vartheta_1 &=& \left(\sigma_1 \otimes \bI_+\right) \otimes \left(\sigma_3 \otimes \bI_-\right) 
 \, , \qquad \sqrt{2} \vartheta_2 = \left(\bI_2 \otimes \bI_+\right) \otimes \left(\sigma_1\otimes \bI_-\right) \, , \nonumber \\
 \xi_n &=& \left(\sigma_2\otimes \chi_n\right) \otimes \left(\sigma_3\otimes \bI_-\right) \, , \qquad \ 
 \tilde\xi_n = \left(\bI_2\otimes \bI_+\right) \otimes \left(\sigma_2 \otimes \tilde\chi_n\right) \, , 
\end{eqnarray}
where $(\chi_n,\chi_n^\dagger)$ and $(\tilde\chi_n,\tilde\chi_n^\dagger)$ are two {\it mutually-commuting}  sets of operators obeying the anticommutation relations 
\be 
\left\{\chi_n, \chi_m^\dagger\right\} =\delta_{nm}\, \bI_+\, , \qquad \left\{\tilde\chi_n,\tilde\chi_m\right\} = \delta_{nm} \, \bI_-\,. 
\ee
In this realization,
\be
\left(-1\right)^{F_R} = \left(\sigma_3\otimes \bI _+\right) \otimes \left(\bI_2 \otimes \bI_- \right) \, , 
\qquad 
\left(-1\right)^{F_L} = \left(\bI_2 \otimes \bI_+\right)\otimes \left(\sigma_3 \otimes\bI_-\right)\, , 
\ee
and hence
\be
(-1)^F= \left(\sigma_3 \otimes \bI_+\right)\otimes \left(\sigma_3 \otimes \bI_- \right)\, . 
\ee

The fermi oscillator ground state is quadruply degenerate; a basis is provided by the four states
\be
 |\varsigma\rangle_+ \otimes  |\tilde\varsigma\rangle_- \qquad \left(\varsigma =\pm \, , \quad \tilde\varsigma= \pm \right), 
\ee
where  
\be
(-1)^{F_R} |\varsigma\rangle_+ = \varsigma |\varsigma\rangle_+\, , \qquad 
(-1)^{F_L} |\tilde\varsigma\rangle_- = \tilde\varsigma |\tilde\varsigma\rangle_-\, . 
\ee
and
\be
\left(\bI_2 \otimes \chi_n\right) |\varsigma\rangle_+= 0\, , \qquad \left(\bI_2\otimes \tilde\chi_n\right) |\tilde\varsigma\rangle_-=0\, . 
\ee
This means that the Fock vacua for the right and left oscillators (bosonic and fermionic) can be chosen to be, respectively, 
\be
|0,\varsigma\rangle_+= |0\rangle_+ \otimes |\varsigma\rangle_+\, , \qquad |0,\tilde\varsigma\rangle_-= |0\rangle_- \otimes |\tilde\varsigma\rangle_-\, , 
\ee
where  $|0\rangle_\pm$ are the Fock vacuum states for the bosonic oscillators, as in (\ref{string-gs}).  The  quadruply-degenerate oscillator  ground state of the string then takes the tensor product form 
\be\label{groundn=2}
|\varsigma,\tilde\varsigma \rangle = |0,\varsigma\rangle_+ \otimes |0,\tilde\varsigma\rangle_- \, . 
\ee

At a given level $L>0$,  the non-hermitian supercharges ${\cal S}_a$ become
\begin{eqnarray}
{\cal S}_1 &=& i\sqrt{2T} L \left[\sigma_1 \otimes \bI_+ 
- i\sigma_2 \otimes \eta_L\right] \otimes \left(\sigma_3\otimes \bI_-\right) \nonumber \\
{\cal S}_2 &=& i\sqrt{2T} L\left(\bI_2 \otimes \bI_+ \right) \otimes \left[\sigma_1 \otimes \bI_- 
- i\sigma_2 \otimes \tilde\eta_L \right]\, ,  
\end{eqnarray}
where both the operators $\eta_L$ and $\tilde\eta_L$, acting in the space of physical states at level $L$, are traceless and square to the identity, and so have (simultaneous) eigenvalues $\pm1$. There are four possible choices of the signs $(\eta_L,\tilde\eta_L)$, and for each choice we get a supermultiplet by the action of ${\cal S}_a^\dagger$ on states annihilated by ${\cal S}_a$. Each such supermultiplet has four states, so there is a minimal 16-fold degeneracy at each non-zero level.

\subsubsection{Low-level excited states}

Excited string states are found, as eigenstates of the level operators $N+\nu$ and $\tilde N+\tilde\nu $ with eigenvalues that we also call $N+ \nu$ and $\tilde N +\tilde\nu$,  by the action of the creation operators on the oscillator vacuum  state such that 
the level-matching condition (\ref{levelmatchn=2}) is satisfied. We may therefore organize all physical  states according to their level $L=N + \nu$, with the corresponding mass being given by (\ref{msquaredn=2}).  Because of the quadruple  degeneracy of the ground state there is a minimal quadruple degeneracy at each level, as required by ${\cal N}=2$ supersymmetry.  There are a total of 16  first excited states, i.e. level-$1$ states:
\begin{eqnarray}\label{l1n2states}
|1_B,\varsigma\rangle_+\otimes|1_B,\tilde\varsigma\rangle_-  &=& \alpha_1^\dagger |0,\varsigma\rangle_+  \otimes \tilde \alpha_1^\dagger |0,\tilde\varsigma\rangle_- , \nonumber \\
  |1_F,\varsigma\rangle_+\otimes|1_F,\tilde\varsigma\rangle_- &=& \chi_1^\dagger |0,\varsigma\rangle_+  \otimes \tilde \chi_1^\dagger |0,\tilde\varsigma\rangle_-  \, , \\ \nonumber
|1_F,\varsigma\rangle_+\otimes|1_B,\tilde\varsigma\rangle_- &=& \chi_1^\dagger |0,\varsigma\rangle_+  \otimes \tilde \alpha_1^\dagger |0,\tilde\varsigma\rangle_- , \nonumber \\
  |1_B,\varsigma\rangle_+\otimes|1_F,\tilde\varsigma\rangle_- &=& \alpha_1^\dagger |0\rangle_+  \otimes \tilde \chi_1^\dagger |0\rangle_-  \, . 
\end{eqnarray}
These form four ${\cal N}=2$ supermultiplets. 

The level-$2$ excited states  are tensor products of the orthonormal states 
\bea\label{l2n2basis-2}
|1_B,1_B,\varsigma\rangle_+  &=& \frac{1}{\sqrt{2}}\left(\alpha_1^\dagger\right)^2  |0,\varsigma\rangle_+ \, , \qquad
 |2_B\rangle_+ =  \frac{1}{\sqrt{2}}\alpha_2^\dagger |0,\varsigma \rangle_+\, ,  \nonumber \\
|1_B,1_F,\varsigma\rangle_+ &=&  \alpha_1^\dagger\chi_1^\dagger |0,\varsigma\rangle_+ \, , \qquad
|2_F,\varsigma\rangle_+ =  \chi_2^\dagger |0,\varsigma\rangle_+\, ,
\eea
with the analogous states built on  $|0,\tilde\varsigma\rangle_-$. This gives us a total of 64 level-$2$ states and hence 16 ${\cal N}=2$ supermultiplets. 

At level $3$ we need to consider the  (orthonormal basis) states
\begin{eqnarray}\label{l3basis-3}
|1_B,1_B,1_B,\varsigma\rangle_+  &=&  \frac{1}{\sqrt{6}} \left(\alpha_1^\dagger\right)^3 |0,\varsigma\rangle_+  ,   \qquad
|1_B,2_B,\varsigma\rangle_+ =   \frac{1}{\sqrt{2}} \alpha_1^\dagger \alpha_2^\dagger |0,\varsigma\rangle_+  , \nonumber \\ 
|3_B,\varsigma\rangle_+  &=&  \frac{1}{\sqrt{3}}\alpha_3^\dagger |0,\varsigma\rangle_+ \, ,  \qquad \qquad 
 |1_F,2_F,\varsigma \rangle_+  =  \chi_1^\dagger \chi_2^\dagger |0,\varsigma\rangle_+  \, , \nonumber \\
  |1_B,2_F,\varsigma\rangle_+  &=&  \alpha_1^\dagger \chi_2^\dagger |0,\varsigma\rangle_+ \, , \qquad \qquad 
   |1_B,1_B,1_F,\varsigma\rangle_+  =  \frac{1}{\sqrt{2}} \left(\alpha_1^\dagger\right)^2\chi_1^\dagger |0,\varsigma\rangle_+ \, , \nonumber \\
 |2_B, 1_F,\varsigma\rangle_+  &=&  \frac{1}{\sqrt{3}}\alpha_2^\dagger\chi_1^\dagger |0,\varsigma\rangle_+ \, ,\qquad \quad
 |3_F,\varsigma\rangle_+ =  \chi_3^\dagger |0,\varsigma\rangle_+ \, .
\end{eqnarray}
Taking tensor products with the corresponding states built on $|0,\tilde\varsigma\rangle_-$ gives a total of 256 states, and hence 64
${\cal N}=2$ supermultiplets. 

To compute the spectrum of superhelicities at these levels we need to compute the eigenvalues of the quantum operator $\varOmega$. In fact, it  is sufficient to compute the  eigenvalues of the operator $\varOmega_+$ because these eigenvalues are identical to those of $\varOmega_-$.   As  neither $\varOmega_+$ nor $\varOmega_-$  depends  on the zero modes $\vartheta_a$, each eigenvalue of $\varOmega$ has at least a  four-fold degeneracy, so  the number  of  eigenvalues of $\varOmega$ at any given level (counting multiplicity)  equals the number of supermultiplets at that level, as required by ${\cal N}=2$ supersymmetry. 

The four ground states are annihilated by $\varOmega$, as must be since (relativistic 3D) superhelicity is not defined for massless particles. These states correspond to massless particles that are potentially identifiable as  a  dilaton and axion,  and their super-partners.  As we already saw for ${\cal N}=0,1$, the operator $\varOmega$ also annihilates the level-$1$ states,  so there are four degenerate copies of the ${\cal N}=2$ supermultiplet of zero superhelicity at this level. The helicity content of this supermultiplet is $s=(-1/2,0,0,1/2)$, so we get  four 3D  ${\cal N}=2$ scalar supermultiplets at level-$1$. 

Similar considerations apply to the higher  levels: the ${\cal N}=2$ helicity content at each level can be deduced directly from the
${\cal N}=1$ results of the previous section. For example, we saw that $\varOmega_+$ has the four eigenvalues $\sqrt{2T}(-\frac{3}{2},-\frac{3}{2},\frac{3}{2},\frac{3}{2})$ at level $2$, so $\varOmega/\sqrt{2T}$ for the ${\cal N}=2$ superstring has eigenvalues $(-3,0,3)$ with multiplicities $(4,8,4)$, leading to 4 ${\cal N}=2$ supermultiplets of superhelicity $s_2=3/2$, another 4 with superhelicity $s_2=-3/2$
and 8 with zero superhelicity (so 16 in total, as required for the 16 supermultiplets at this level). The $s_2=3/2$ supermultiplet has helicities $s=(2,3/2,3/2,1)$;  it is a massive spin-2 ${\cal N}=2$ supermultiplet. 

As these results show, the states of the ${\cal N}=2$ 3D superstring through level $2$ are just standard bosons and fermions, but this simple feature does not extend to level $3$. Again, the level-$3$ content can be deduced from the previous results for  ${\cal N}=1$. 
From the eight level-$3$ eigenvalues of $\varOmega_+$ given in (\ref{Omega3}) we get a total of 64 eigenvalues of $\varOmega$, as required for the 64 supermultiplets at this level.  Eight of them have zero superhelicity but the rest have irrational superhelicities. As all helicities in such a  supermultiplet  are also irrational, we conclude not only that there are  anyons in the spectrum of the ${\cal N}=2$ 3D superstring, but also that these anyons are  `generic'  ones of irrational spin.


\section{Summary and Outlook}

The quantum theory of strings below their  critical dimension is problematic and generically involves the introduction of a new degree of freedom, the Liouville mode.  We say ``generically'' because there is an exception: the usual quantum Lorentz anomaly in the light-cone gauge, in which the action involves only physical worldvolume variables, is trivially absent for the Nambu-Goto string in a Minkowski spacetime of three dimensions (3D) \cite{Mezincescu:2010yp, Curtright:2010zz}, so 
no Liouville mode is needed to guarantee unitarity and Lorentz invariance, at least for a free 3D Nambu-Goto string.  

The implication is that the quantum spectrum contains states of definite mass and spin, and this was verified explicitly in  \cite{Mezincescu:2010yp}, with a rather surprising result: the spins are not generically integer or half-integer.  This is possible  because the rotation subgroup of the universal cover $\overline{SO}(1,2)$ of the 3D Lorentz group is $\overline{SO}(2)\cong \bR$. In the context  of a relativistic theory, this implies that the states in the string spectrum generically describe ``anyons''.  There is an ambiguity in the string spectrum due to an operator ordering ambiguity: the mass-squared  of  the string ground state is arbitrary, although  it must be non-negative to avoid tachyons. This ambiguity  affects the spins as well as the masses.
Consideration of both the level-$2$ and level-$3$ excited states led to the conclusion that some states are necessarily anyons and that they generically have irrational spin.

A similar conclusion was arrived at for  the ${\cal N}=1$ 3D Green-Schwarz (GS) superstring, but in that case  the doubly-degenerate ground state is required by supersymmetry to be massless, 
so the quantum ambiguity of the bosonic 3D string is eliminated.  The level-$2$ and level-$3$ excited states were shown in  \cite{Mezincescu:2010yp} to contain  ``semions'' (a particular case of anyons for which the spin is $1/4$ modulo a half-integer). In this paper we have given details of the computations behind these results,  and we have extended them in a number of ways.  

{}Firstly, we have extended the computation of the spectrum of the quantum 3D Nambu-Goto string to level-$4$. This allows us to strengthen our earlier  conclusion concerning anyons in the spectrum: some of these anyons necessarily have {\it irrational} spin. This tells us that the Lorentz group really is $\overline{SO}(1,2)$ and not some finite cover of $SO(1,2)$.  
Secondly, we have  established the same result for the ${\cal N}=1$ superstring by showing that irrational spin anyons are present in the spectrum at level-$3$.  We have also established
the absence of super-Poincar\'e anomalies. Classically, there are actually two ${\cal N}=1$
superstring theories, interchanged by worldsheet parity, because the string fermions propagate in one direction around the string. However, these two equivalent, but distinct, classical theories are identical as quantum theories because they describe exactly the same 3D spectrum. 

Thirdly, and this is our main new result, we have extended the analysis to include the ${\cal N}=2$ GS superstring. In this case, the spectrum  through level $2$ consists only of bosons and fermions (i.e. particles of integer and half-odd-integer spins) so it was not previously clear to us whether the spectrum would contain anyons. In fact, the level-$3$ spectrum contains  particles of irrational spin, this being a consequence of  the presence of such states in the ${\cal N}=1$ superstring. 

The fact that irrational spins appear in the spectrum  of all 3D (super)strings implies that the Lorentz group is the infinite universal cover of $SO(1,2)$, not the double cover that might have been expected, nor any finite multiple cover. We believe that this may explain why existing covariant quantization methods do not appear to allow for the possibility of 3D strings: covariant quantization of even a free 3D  {\it particle} is not straightforward if it has irrational spin. 

We have made no attempt to explore whether the free 3D strings discussed here admit interactions. Again, this  is already a difficult problem for particles of irrational spin. If interactions are possible then one would expect there to exist effective supersymmetric field theories describing the massless modes of the ${\cal N}=1$ and ${\cal N}=2$ 3D superstring theories. Our results are consistent with  this possibility even if the effective field theories are supposed to be supergravity theories because neither the metric nor the antisymmetric tensor fields that couple naturally to a string propagate massless modes in 3D. 

{}For  the ${\cal N}=2$ superstring, there are four massless  states:  a scalar and a pseudo-scalar,  and their superpartners. The scalar might be interpretable as a dilaton. As a  massless pseudo-scalar is dual to a massless vector in 3D,  it would be natural to suppose that any effective field theory is some generally covariant theory involving an ${\cal N}=2$ vector multiplet. The vector potential of this supermultiplet could  couple to particles carrying the  central charge permitted by the  ${\cal N}=2$ superalgebra. Although there are no such particles in the spectrum of a free ${\cal N}=2$ superstring, they might be non-perturbative excitations of an interacting ${\cal N}=2$ 3D superstring, analogous to the D0-branes of  critical superstring theory. If so, they might show up in an analysis of ${\cal N}=2$ open strings with Dirichlet boundary conditions.

{}Finally, we recall that the  ${\cal N}=2$ 3D GS superstring is, classically,  the double-dimensional reduction of the 4D supermembrane. In the context of a  4D   spacetime that is a product of  3D Minkowski spacetime with a circle, the supermembrane can be wrapped on the circle to give a string. The ${\cal N}=2$ 3D superstring is then found by  ignoring the momentum modes in the extra dimension, but it would be interesting to see what effect these modes have on the string spectrum, and whether there are other implications of a 4D perspective.

\subsection*{Acknowledgments}
We thank Michael Green for a helpful discussion.  LM acknowleges partial support from National Science Foundation Award 0855386.

\end{document}